\documentclass[longauth]{aa} 
\usepackage{txfonts}
\usepackage{graphicx}
\usepackage{placeins}
\usepackage{rotating}
\usepackage{subcaption}
\usepackage[breaklinks=true]{hyperref}
\usepackage{natbib,twoopt}
\bibpunct{(}{)}{;}{a}{}{,}           
\makeatletter
  \newcommandtwoopt{\citeads}[3][][]{\href{http://adsabs.harvard.edu/abs/#3}%
    {\def\hyper@linkstart##1##2{}%
     \let\hyper@linkend\@empty\citealp[#1][#2]{#3}}}
  \newcommandtwoopt{\citepads}[3][][]{\href{http://adsabs.harvard.edu/abs/#3}%
    {\def\hyper@linkstart##1##2{}%
     \let\hyper@linkend\@empty\citep[#1][#2]{#3}}}
  \newcommandtwoopt{\citetads}[3][][]{\href{http://adsabs.harvard.edu/abs/#3}%
    {\def\hyper@linkstart##1##2{}%
     \let\hyper@linkend\@empty\citet[#1][#2]{#3}}}
  \newcommandtwoopt{\citeyearads}[3][][]%
    {\href{http://adsabs.harvard.edu/abs/#3}
    {\def\hyper@linkstart##1##2{}%
     \let\hyper@linkend\@empty\citeyear[#1][#2]{#3}}}
\makeatother
\usepackage{fancyvrb}
\usepackage{listings}
\RecustomVerbatimCommand{\VerbatimInput}{VerbatimInput}%
{fontsize=\footnotesize,
 frame=lines,  
 framesep=2em, 
 rulecolor=\color{Gray},
 labelposition=topline,
 commandchars=\|\(\), 
 commentchar=*        
}

\begin{document} 

   \title{J-PLUS: The stellar mass function of quiescent and star-forming galaxies at $0.05 \leq {\rm z} \leq 0.2$}

   \author{F.~D.~Arizo-Borillo\inst{1}
          \and
          C.~López-Sanjuan\inst{1,2}
          \and 
          I.~Pintos-Castro\inst{1}
          \and
          J.~A.~Fernández-Ontiveros\inst{1,2}
          \and 
          T.~Kuutma\inst{1}
          \and
          A.~Lumbreras-Calle\inst{1}
          \and
          A.~Hernán-Caballero \inst{1,2}
          \and 
          H.~Domínguez-Sánchez \inst{3,1}
          \and 
          G.~De Lucia \inst{4,5}
          \and 
          F.~Fontanot \inst{4,5}
          \and
          L.~A. Díaz-García\inst{6}  
          \and J. M. Vílchez\inst{6}
          \and P. T. Rahna\inst{1}
          \and A.~J.~Cenarro\inst{1,2}
          \and D.~Crist\'obal-Hornillos\inst{1}
          \and C.~Hern\'andez-Monteagudo\inst{7,8}  
          \and A.~Mar\'{\i}n-Franch\inst{1,2}
          \and M.~Moles\inst{1}
          \and J.~Varela\inst{1}
          \and H.~V\'azquez Rami\'o\inst{1,2}
          \and J.~Alcaniz\inst{9}
          \and R.~A.~Dupke\inst{9,10}
          \and A.~Ederoclite\inst{1,2}
          \and L.~Sodr\'e Jr.\inst{11}
          \and R.~E.~Angulo\inst{12,13}
          }

   \institute{Centro de Estudios de F\'{\i}sica del Cosmos de Aragón (CEFCA), Plaza San Juan 1, 44001 Teruel, Spain\label{CEFCA}\\
        \email{farizo@cefca.es, franarizob@gmail.com}
         \and Unidad Asociada CEFCA-IAA, CEFCA, Unidad Asociada al CSIC por el IAA y el IFCA, Plaza San Juan 1, 44001 Teruel, Spain
         \and Instituto de Física de Cantabria-CSIC, Avda. de los Castros s/n 39005, Santander, Spain
         \and INAF - Astronomical Observatory of Trieste, via G.B. Tiepolo 11, I-34143 Trieste, Italy
         \and IFPU - Institute for Fundamental Physics of the Universe, via Beirut 2, 34151, Trieste, Italy
         \and Instituto de Astrofísica de Andalucía (CSIC), PO Box 3004, 18080 Granada, Spain
         \and Instituto de Astrof\'{\i}sica de Canarias, La Laguna, 38205, Tenerife, Spain
         \and Departamento de Astrof\'{\i}sica, Universidad de La Laguna, 38206, Tenerife, Spain
         \and Observat\'orio Nacional - MCTI (ON), Rua Gal. Jos\'e Cristino 77, S\~ao Crist\'ov\~ao, 20921-400 Rio de Janeiro, Brazil
         \and University of Michigan, Department of Astronomy, 1085 South University Ave., Ann Arbor, MI 48109, USA
         \and Instituto de Astronomia, Geof\'{\i}sica e Ci\^encias Atmosf\'ericas, Universidade de S\~ao Paulo, 05508-090 S\~ao Paulo, Brazil
         \and Donostia International Physics Centre (DIPC), Paseo Manuel de Lardizabal 4, 20018 Donostia-San Sebastián, Spain
         \and IKERBASQUE, Basque Foundation for Science, 48013, Bilbao, Spain
             }

   \date{Received ?? , 2025; accepted ??}

\abstract
{}
{We derive the stellar mass function (SMF) of quiescent and star-forming galaxies at ${\rm z} \leq 0.2$ using 12-band (five broad plus seven narrow) photometry from the Javalambre Photometric Local Universe Survey (J-PLUS) third data release (DR3) over $3\,284$ deg$^2$, where the narrow bands improve photometric-redshift and stellar property precision relative to purely broadband surveys.
}
{We selected $\sim 890\,000$ galaxies with $r \leq 20$ mag and photometric redshifts in the range $0.05 \leq z \leq 0.20$ over an effective area of $2\,881$ deg$^2$, corresponding to a comoving volume of $V \simeq 1.6 \times 10^{8}\,\mathrm{Mpc}^3$. Stellar masses and star formation rates were derived through spectral energy distribution (SED) fitting with the Code Investigating GALaxy Emission (\texttt{CIGALE}), and confronted with spectroscopic samples. Galaxies were classified as star-forming or quiescent based on their specific star formation rate (sSFR), adopting a threshold of $\log\ (\mathrm{sSFR}\ [\mathrm{yr}^{-1}]) = -10.2$. We computed SMFs for both populations using the $1/V_{\mathrm{max}}$ method, applied completeness corrections, and fit a parametric single Schechter function.
}
{The SMFs derived from J-PLUS DR3 are well described by Schechter functions and agree with previous photometric and spectroscopic studies. The characteristic mass for quiescent galaxies, $\log\ (M_{\star}/M_{\odot}) = 10.80$, is higher by $0.4$ dex than that of star-forming galaxies. The faint-end slope is steeper for star-forming galaxies ($\alpha = -1.2$) than for quiescent ones ($\alpha = -0.7$). The quiescent fraction increases by $40$\% per dex in stellar mass, reaching $f_{\rm Q} > 0.95$ at $\log\ (M_{\star}/M_{\odot}) > 11$. Comparisons with the GAlaxy Evolution and Assembly (GAEA) semi-analytic model show an overabundance of simulated star-forming galaxies, particularly at intermediate masses.}
{The SMFs and quiescent fraction from J-PLUS DR3 are consistent with the literature and provide valuable constraints for galaxy formation models. Quiescent galaxies represent $45$\% of the number density at $\log M_{\star} > 9$ but contribute $75$\% of the stellar mass density. This work lays the groundwork for studies of environmental quenching using J-PLUS. The inclusion of seven narrowband filters improves redshift precision by $20$\%, enabling more accurate SED fitting and galaxy classification. These methods and findings can be extended with J-PAS, which will provide deeper and higher-resolution photometry over a wider spectral range.}


\maketitle
\nolinenumbers
\section{Introduction}
The stellar mass function (SMF) is a fundamental tool for understanding the properties and evolution of galaxy populations, encompassing both quiescent (Q) and star-forming (SF) systems. It represents the number density of galaxies as a function of their stellar mass, $M_{\star}$, determined by counting galaxies in stellar mass bins in a given cosmological volume and correcting from biases. The SMF provides critical insights into the evolution of the star formation rate (SFR) over cosmic time, enabling us to track the accumulation of stellar mass across different epochs. Large redshift surveys, such as the Sloan Digital Sky Survey SDSS \citep{york2000sloan}, zCOSMOS \citep{lilly2007zcosmos}, and the Galaxy And Mass Assembly (GAMA) survey \citep{baldry2010galaxy}, have significantly advanced our ability to study galaxy populations with unprecedented precision. These surveys facilitate the derivation of the SMF and its evolution across cosmic time, substantially contributing to our understanding of galaxy formation and evolution (see also \citealt{muzzin2013evolution}, \citealt{peng2010mass}, and \citealt{kelvin2014galaxy}). More recent studies, including those from GAMA \citep{wright2018gama,driver2022galaxy}, have further refined and extended these analyses.

The SMF is not only pivotal for observational studies but also serves as a crucial constraint for theoretical models of galaxy formation. In these models, the SMF works as a key diagnostic tool for testing model predictions and refining simulation parameters. A primary challenge in these models is reconciling the halo mass function from simulations with the observed SMF from galaxy surveys. Addressing discrepancies, especially at the low- and high-mass ends, requires iterative refinements through feedback mechanisms, such as supernovae and active galactic nuclei (AGN) feedback, which regulate gas cooling and align the predicted most massive end of the SMF with observed data \citep{kauffmann1993formation, somerville2008semi}. Up to $z \sim 7.5$, the SMF is well described by a single or double Schechter function \citep{schechter1976analytic,weaver2023cosmos2020}, featuring an exponential decline at a characteristic stellar mass ($M^*$) and a low-mass slope ($\alpha$). While the low-mass slope shows only mild evolution, several studies report a moderate but significant increase in $M^*$ from $z \sim 2$ to the present day \citep{marchesini2009evolution, tomczak2014galaxy, adams2021evolution}, likely driven by mass growth through mergers. This suggests that the mechanisms regulating stellar mass assembly vary across cosmic time. In contrast, the normalization of the SMF ($\Phi^*$), especially for SF galaxies, evolves more strongly and reflects the decline in cosmic SFR density \citep{popesso2023main}. Further discussion of these trends can be found in \citet{weaver2023cosmos2020, wright2018gama}.

Despite these advancements, the underlying physical mechanisms governing the SMF, even in the low-redshift Universe, remain poorly understood. Processes such as galaxy-galaxy mergers, stellar and gas kinematics, gas inflows and outflows, and feedback from supernovae and massive black holes are active areas of research. These mechanisms play a crucial role in the transformation of SF galaxies into passive ones, making them a subject of intense investigation in contemporary astrophysics. Studying the SMF separately for SF and Q galaxies helps reveal the physical processes that quench star formation, which are not used to calibrate theoretical models but provide valuable tests of their predictions. Two primary mechanisms are frequently invoked to explain the quenching of star formation: AGN feedback and environmental effects. In the case of AGN feedback, the accretion of matter by a central supermassive black hole injects energy and momentum into the surrounding medium, heating the gas and suppressing star formation. Environmental effects, particularly in high-density regions such as galaxy clusters, can also lead to quenching. Mechanisms such as gas stripping and galaxy interactions can inhibit star formation. While AGN feedback is most efficient in massive galaxies ($M_{\star} > 10^{10.5}\ M_{\odot}$) \citep{Piotrowska2022}, environmental factors primarily influence dwarf galaxies ($M_{\star} < 10^9\ M_{\odot}$; \citealt{peng2010mass}).

In this study, we used data from the Javalambre Photometric Local Universe Survey (J-PLUS\footnote{\href{j-plus.es}{j-plus.es}}; \citealt{cenarro2019j}) third data release (DR3) to estimate the SMF in the local Universe. J-PLUS DR3 covers a sky area of $3\,284$ square degrees and employs a filter set comprising the five broad bands ($ugriz$) and seven narrow bands, thereby providing improved photometric redshifts (${\rm z}_{\rm phot}$) and SFRs with respect to previous broadband surveys. A sample of $890\,844$ galaxies at redshift $0.05 \leq {\rm z} \leq 0.20$ and with $r \leq 20$ mag was used to estimate the SMF for both SF and Q galaxies. Such a large sample size is critical for conducting robust statistical studies of galaxy populations.
    
This paper is structured as follows. Section~\ref{sec:data} describes the dataset and sample selection. Section 3 outlines the methodology, including quality cuts, spectral energy distribution (SED) fitting, and the 1/$V_{\text{max}}$ method for deriving the SMF in J-PLUS DR3 for Q and SF galaxies. Section 4 presents the results, including the SMF for Q, SF, and all galaxies, along with the Q fraction. The discussion and conclusions are provided in Sects.~\ref{sec:Discussion} and \ref{sec:Conclusions}. Throughout this work, we adopt a Lambda cold dark matter ($\Lambda$CDM) cosmology with $H_{0} = 70 \ \text{km} \ \text{s}^{-1} \ \text{Mpc}^{-1}$, $h = 0.7$, $\Omega_{m} = 0.3$, and $\Omega_{\Lambda} = 0.7$.

\section{Data and galaxy properties}\label{sec:data}
\subsection{J-PLUS DR3}
J-PLUS is carried out at the Observatorio Astrofísico de Javalambre (OAJ\footnote{\href{https://oajweb.cefca.es/}{https://oajweb.cefca.es/}};\citealt{cenarro2014observatorio}) using the 83 cm Javalambre Auxiliary Survey Telescope (JAST80). The telescope is equipped with a 9.2k x 9.2k pixel camera \citep{t80cam}, providing a wide field of view of 2 deg$^2$. It features a set of $12$ photometric filters, including the five SDSS ($u$,$g$,$r$,$i$,$z$) broadband filters and seven medium or narrowband filters. These additional filters target key stellar spectral features: four cover the region around the 4\,000\AA \ break ($J0378$, $J0395$, $J0410$, and $J0430$), one captures the magnesium doublet ($J0515$), another probes the calcium triplet ($J0861$), and another is centered on the H$\alpha$ emission line at rest frame ($J0660$). These narrow band filters allow more accurate estimates of photometric redshifts and galaxy physical properties. Further details about the J-PLUS observing strategy, data reduction, and general goals can be found in \citep{cenarro2019j}. Consequently, J-PLUS provides broad optical coverage, enabling a variety of studies in stellar astrophysics \citep[e.g.,][]{bonatto2019j, whitten2019j, solano2019j, lopez2024}, galaxy evolution across different redshift ranges \citep[e.g.,][]{logrono2019j, san2019j}, and galaxy clusters \citep[e.g.,][]{molino2019j, jimenez2019j}. Additionally, it has been used to identify extreme extragalactic emitters \citep[e.g.,][]{spinoso2020j, lumbreras2022}.

For this study, we utilized J-PLUS DR3, which was made publicly available in December $2022$. J-PLUS DR3 includes photometric information for $1\,642$ pointings, covering a total sky area of $3\,284$ deg$^2$. This was reduced to $2\,881$ deg$^2$ after masking bright stars, optical artifacts, and overlapping regions. The catalog and different tables used here are publicly accessible on the J-PLUS website\footnote{\href{https://www.j-plus.es/datareleases/data_release_dr3}{https://www.j-plus.es/datareleases/data\_release\_dr3}}, which contains calibrated photometry \citep{lopez2023j} in several apertures for approximately $47.4$ million objects detected in the $r$ band, obtained using SExtractor's dual-mode.

\begin{figure*}[ht]
    \centering
        \includegraphics[width=0.75\textwidth]{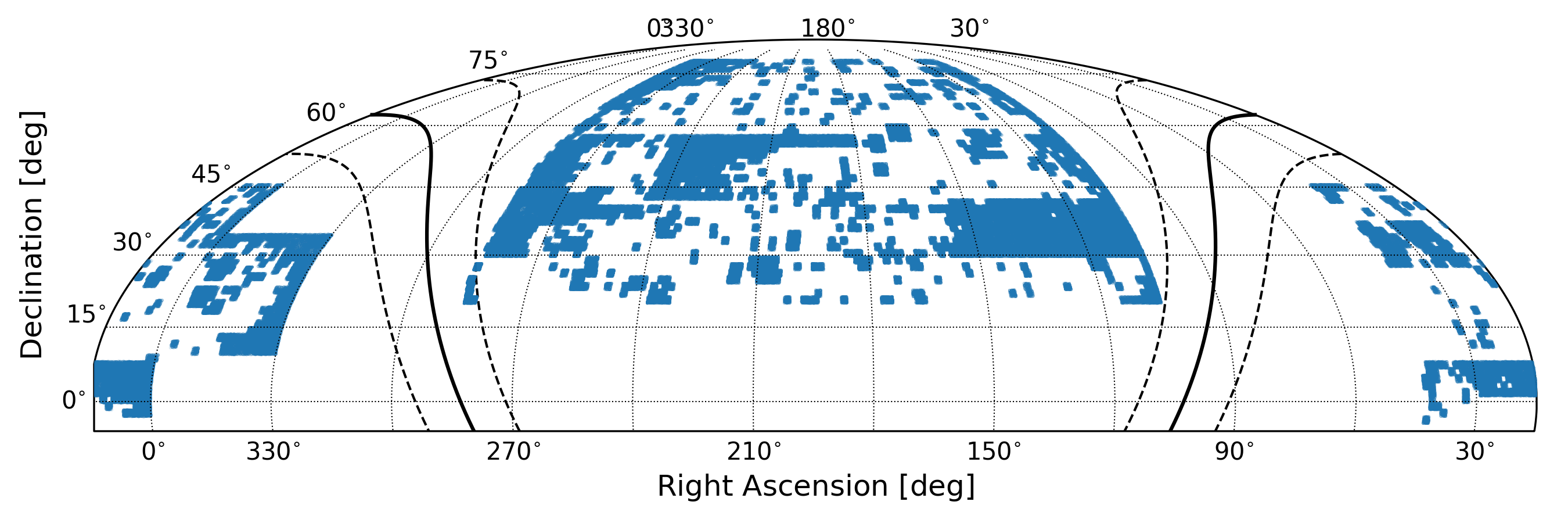}
        \caption{Sky location of J-PLUS DR3 galaxies with $r \leq 20$ mag at $0.05 \leq {\rm z}_{\rm phot} \leq 0.20$ covering $2\,881$ deg$^2$ (blue points). The black curve depicts the Milky Way.}
        \label{fig:footprintdr3}
\end{figure*}

\subsection{J-PLUS DR3 galaxy sample}\label{sec:sample}
To select galaxies, we used the point-like probability from Table \texttt{galclass.sglc\_prob\_star} in \citep{lopez2019j}, estimated by combining the available priors and information in the $gri$ broad bands. Sources with probabilities lower than $0.5$ were selected as galaxies.

From the \texttt{jplus.MagABDualObj} table, we retrieved the magnitudes extracted using AUTO aperture (elliptical apertures with semimajor axis equal to twice the Kron radius of the source). The magnitudes of J-PLUS DR3 galaxies were corrected for Galactic extinction using the \texttt{jplus.MWExtinction} table, which contains $B-V$ color excess due to Milky Way dust at the position source. The integrated $E(B-V)$ was estimated from the \cite{schlafly2011measuring} recalibration of the \cite{schlegel1998maps} infrared-based dust map. 
The lower redshift limit $z = 0.05$ was chosen to minimize the effects of aperture and surface-brightness and to reduce the impact of large-scale local structures on the measured SMF. The photometric redshifts are available in the \texttt{jplus.PhotoZLephare} table. These redshifts were obtained using \texttt{LePhare} \citep{arnouts2011lephare}. \texttt{LePhare} estimates the photometric redshifts in J-PLUS DR3 by scanning from ${\rm z}_{\rm min} = 0.0$ to ${\rm z}_{\rm max} = 1.0$ in steps of ${\rm z}_{\rm step} = 0.005$. At each redshift, it computes a likelihood based on the $\chi^2$ of the best-fitting template to the J-PLUS data. The CEFCA\_minijpas library used includes $50$ synthetic galaxy spectra produced with the Code Investigating GALaxy Emission (\texttt{CIGALE}; \citealt{boquien2019cigale}). A prior derived from the VIMOS VLT Deep Survey (VVDS) galaxy counts \citep{le2005vimos} was applied to obtain the final redshift probability distribution. For details on the configuration, templates, and priors, we refer to \citet{hernan2021minijpas}.

The ADQL query used to access the J-PLUS DR3 data used in this work is presented in Appendix~\ref{app:adql}, with a total of $890\,844$ galaxies with extinction-corrected $r_{0} \leq 20$ mag at $0.05 \leq {\rm z}_{\rm phot} \leq 0.20$ and $\texttt{MASK\_FLAGS} = 0$ (Fig.~\ref{fig:footprintdr3}). The selected redshift and magnitude ranges ensured enough signal-to-noise in the J-PLUS photometry and a proper covering of the $4\,000\ \AA$ break with the bluest medium-band filters to derive high-quality physical properties for the galaxies (Sect.~\ref{sec:cigale}).

The accuracy of the $z_{\rm phot}$ in the sample was estimated by comparing them with the spectroscopic redshifts (${\rm z}_{\rm spec}$) from SDSS DR12, which has $686\,176$ counterparts. The difference
\begin{equation}
    \delta {\rm z} = \dfrac{{\rm z}_{\rm phot}-{\rm z}_{\rm spec}}{1+{\rm z}_{\rm spec}} 
\end{equation}
was computed. A negligible systematic offset of $-0.002$ and a $0.011$ dispersion were measured from a Gaussian fit to the $\delta {z}$ distribution. For comparison, the Gaussian distribution compared to the photometric redshifts of \cite{beck2016photometric} based on $ugriz$ SDSS photometry yields a dispersion of $0.013$, i.e., a $20$\% improvement when J-PLUS DR3 photometry is used.

\subsection{Physical properties of galaxies with \texttt{CIGALE}}\label{sec:cigale}
Spectral energy distribution modeling was performed using \texttt{CIGALE}, with the full configuration provided in Appendix~\ref{app:cigale}. For each galaxy, \texttt{CIGALE} was run at a fixed redshift, adopting the \texttt{LePhare} photometric redshift as an input. The star formation history was modeled using the \texttt{sfhdelayed} module, adopting a delayed-$\tau$ model with optional recent bursts. The main stellar population spans a range of three $\tau_\mathrm{main}$ values from $100$ to $5\,000$~Myr and five ages from $3000$~Myr to $13$~Gyr, while the burst component includes three $\tau_\mathrm{burst}$ values from $5$ to $500$~Myr, six ages  between $5$ and $1000$~Myr, and four burst mass fractions up to $40$\%. The spectra of stellar populations were synthesized using the \citet{BC03} models with a \citet{Chabrier2003} initial mass function (IMF) and three metallicities ($Z = 0.004$, $0.008$, and $0.02$). Nebular emission was included using the \texttt{nebular} module, with ionization parameters $\log U = -2.0$ and $-3.5$, gas-phase metallicities $Z_\mathrm{gas} = 0.014$ and $0.022$, electron density $n_e = 100~\mathrm{cm}^{-3}$, and ionizing photon escape fractions up to 20\%. Dust attenuation was modeled with the \texttt{dustatt\_modified\_starburst} module, following a Calzetti-like law \citep{Calzetti2000} with five $E(B-V)$ values up to $0.5$, no UV bump, and a power-law slope of $0.0$ ($R_V = 3.1$).

\begin{figure*}[htpb!]
\begin{subfigure}{.45\linewidth}
        \includegraphics[width=\linewidth]{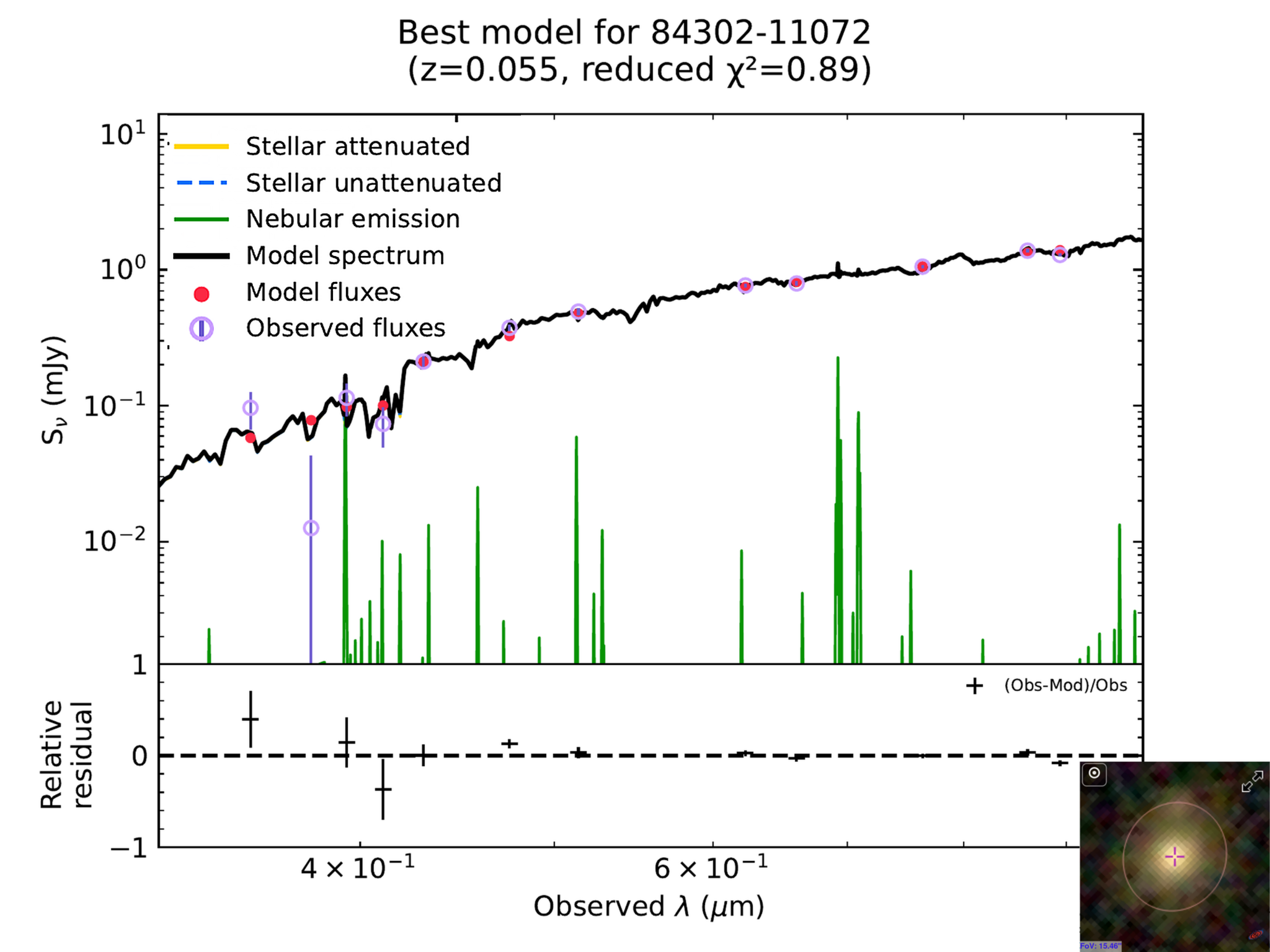}
        \caption{}
        \label{fig:cigale_q_fit}
\end{subfigure}\hfill 
\begin{subfigure}{.45\linewidth}
        \includegraphics[width=\linewidth]{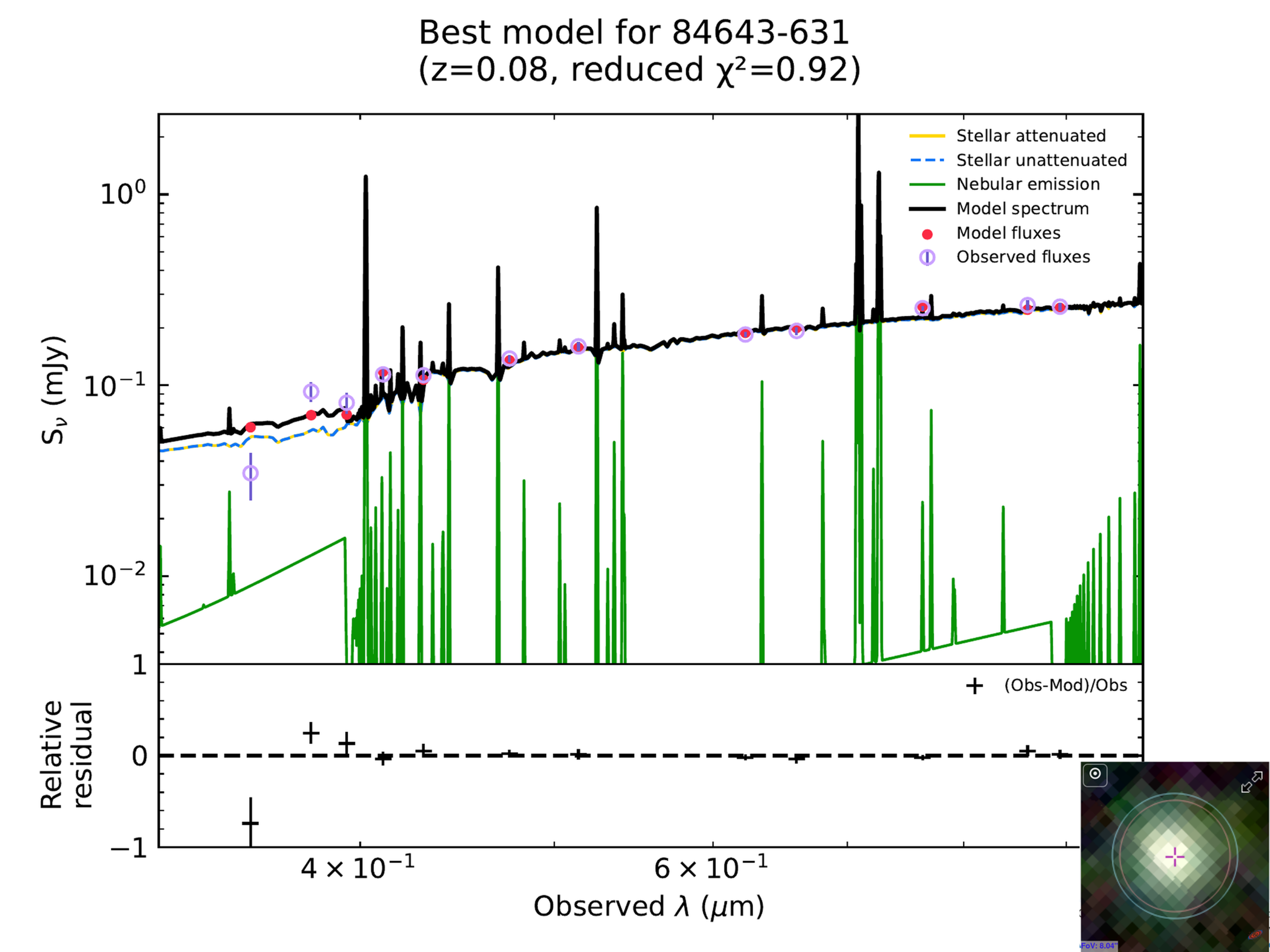}
        \caption{}
        \label{fig:cigale_sf_fit}
\end{subfigure}
\caption{Examples of SED fitting to J-PLUS 12-band photometry using \texttt{CIGALE}. Panel~(a): Q galaxy. Panel~(b): SF galaxy. In each panel, the observed fluxes are shown as purple circles with error bars, and the best-fit model spectrum is plotted as a solid black line. The red dots indicate the model-predicted fluxes in each filter. The contributions from the attenuated stellar population (orange line), unattenuated stellar emission (dashed blue line), and nebular emission (green line) are also shown, although difficult to see in the left panel due to their resemblance to the model spectrum. Bottom subpanels: Relative residuals, defined as $(\mathrm{Obs} - \mathrm{Mod})/\mathrm{Obs}$, with a horizontal dashed line at zero for reference. A $gri$ composite image from J-PLUS is inset in the bottom right corner of each panel.}

\label{fig:cigale_sed_fit}
\end{figure*}

The derived parameters used in this study are the stellar mass of the galaxy (in $M_{\odot}$ units) and its SFR averaged over the last $10$ Myr, in $M_{\odot}\,{\rm yr}^{-1}$ units. This SFR was used to cover the typical visibility timescale of the widely used H${\alpha}$ emission line of a star-formation burst. The specific SFR for each galaxy was defined as ${\rm sSFR} = {\rm SFR} / M_{\star}$ [yr$^{-1}$]. Two examples are presented in Fig.~\ref{fig:cigale_sed_fit}. We observe that both fittings are successful, with reduced $\chi^2 \approx 1$. The table for the nearly $900$k galaxies analyzed is available in the J-PLUS DR3 repository and in VizieR.

The reliability of the derived galaxy properties was tested by comparison with the results from \citet{puertas2017aperture}. They estimated the stellar masses, SFRs, and specific star formation rates (sSFRs) for a sample of $209\,276$ SF galaxies with spectroscopic information from SDSS. An empirical correction based on the Calar Alto Legacy Integral Field Area (CALIFA; \citealt{sanchez2012califa}) spectroscopic data to the fiber measurements in SDSS was applied, providing an excellent reference for our photometric measurements. The difference between the \citet{puertas2017aperture} and J-PLUS measurements were defined as
\begin{equation}
    \delta \mathcal{X} = \log \mathcal{X}^{\rm DP} - \log \mathcal{X}^{\rm J-PLUS},
\end{equation}
where the parameters $\mathcal{X} = \{M_{\star}, {\rm SFR}, {\rm sSFR}\}$ were explored. The distribution of the differences for the $14\,063$ objects in common was approximated with a Gaussian distribution. The median ($\mu$) and the dispersion ($\sigma$) of the Gaussian were used to determine the quality of the J-PLUS measurements. 

\begin{figure*}[htpb!]
\begin{subfigure}{.33\linewidth}
        \includegraphics[width=\linewidth]{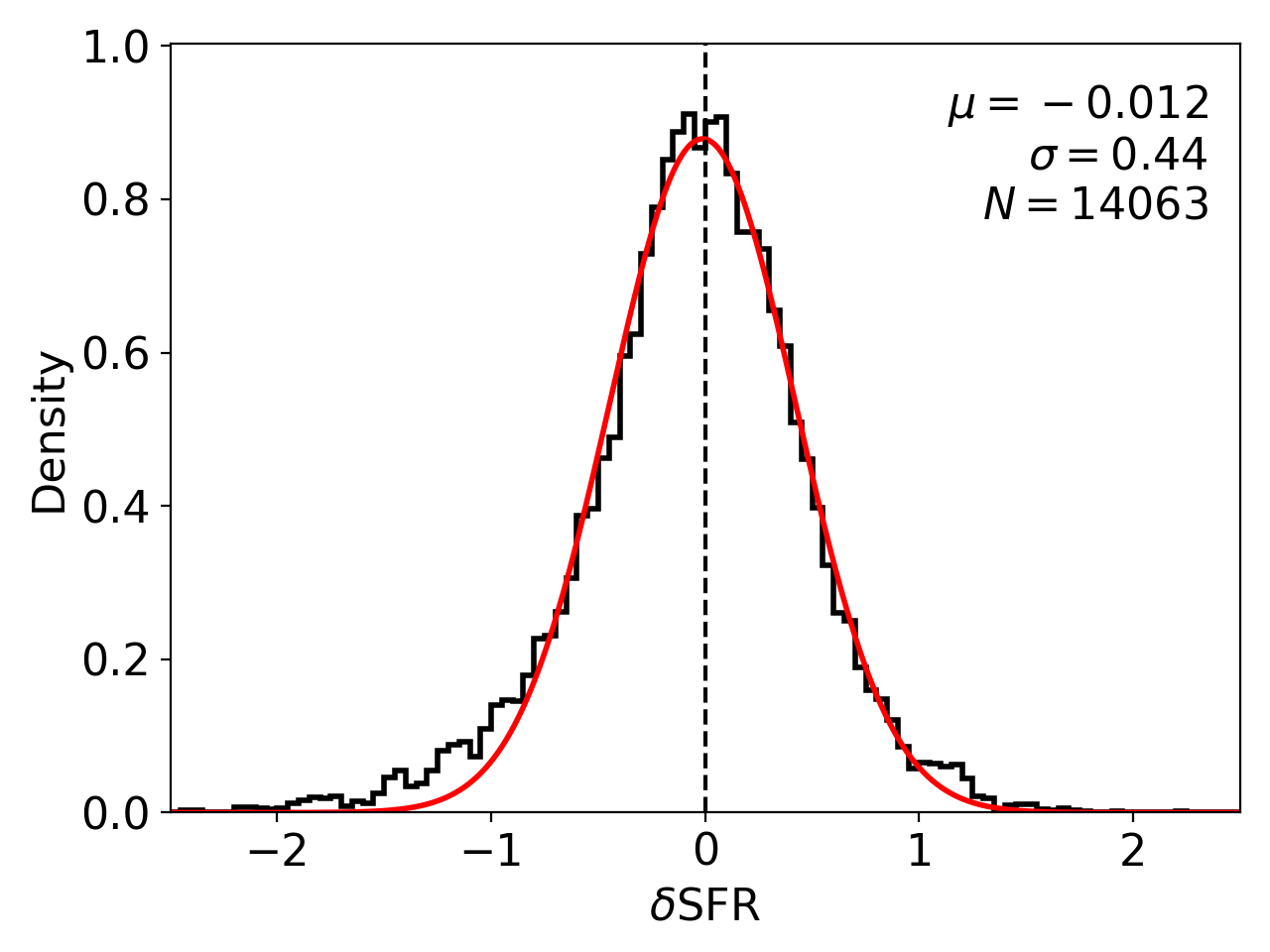}
        \caption{}
        \label{fig:sfr_puertas}
\end{subfigure}\hfill
\begin{subfigure}{.33\linewidth}
        \includegraphics[width=\linewidth]{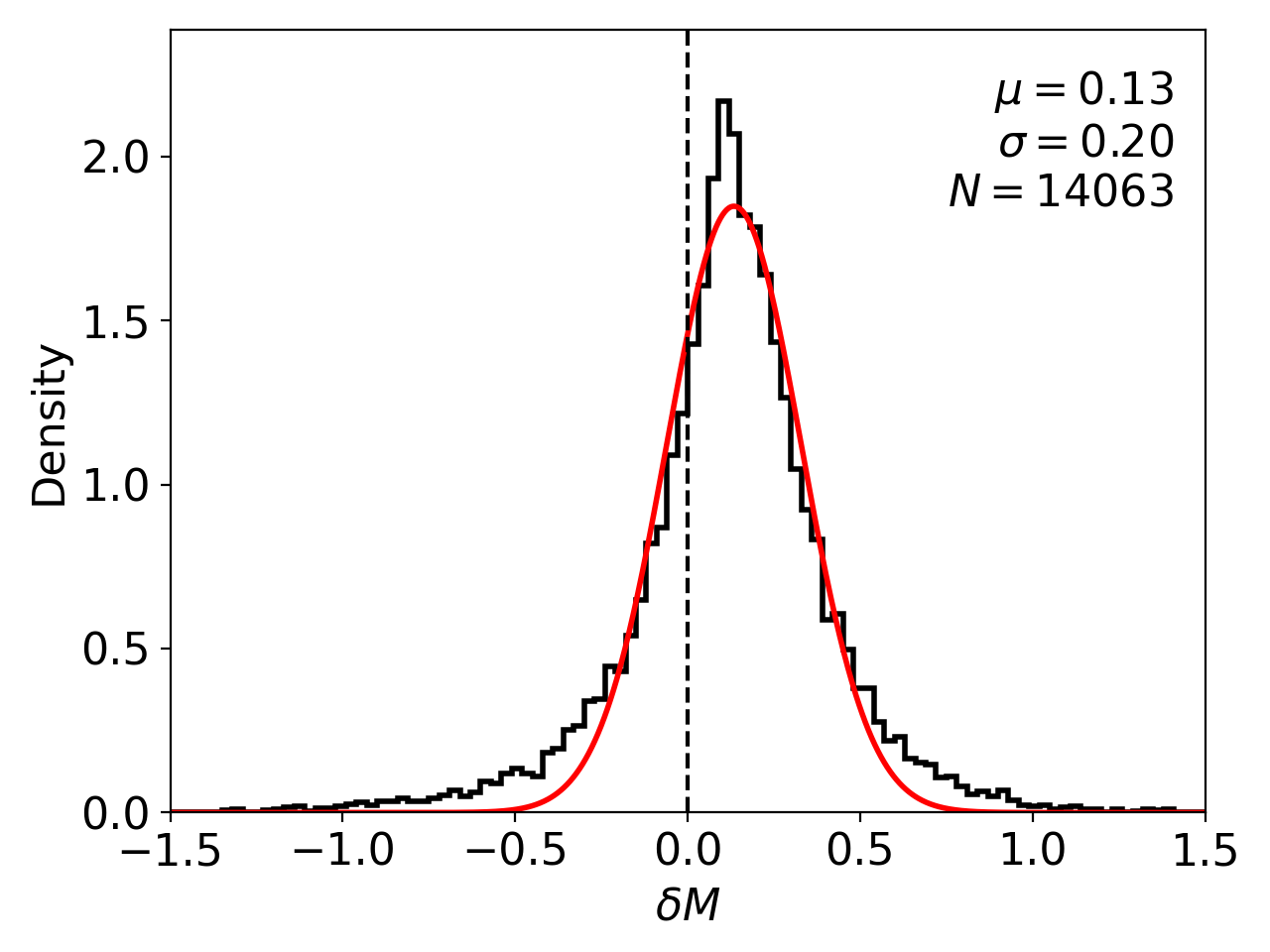}
        \caption{}
        \label{fig:mass_puertas}
\end{subfigure}
\begin{subfigure}{.33\linewidth}
        \includegraphics[width=\linewidth]{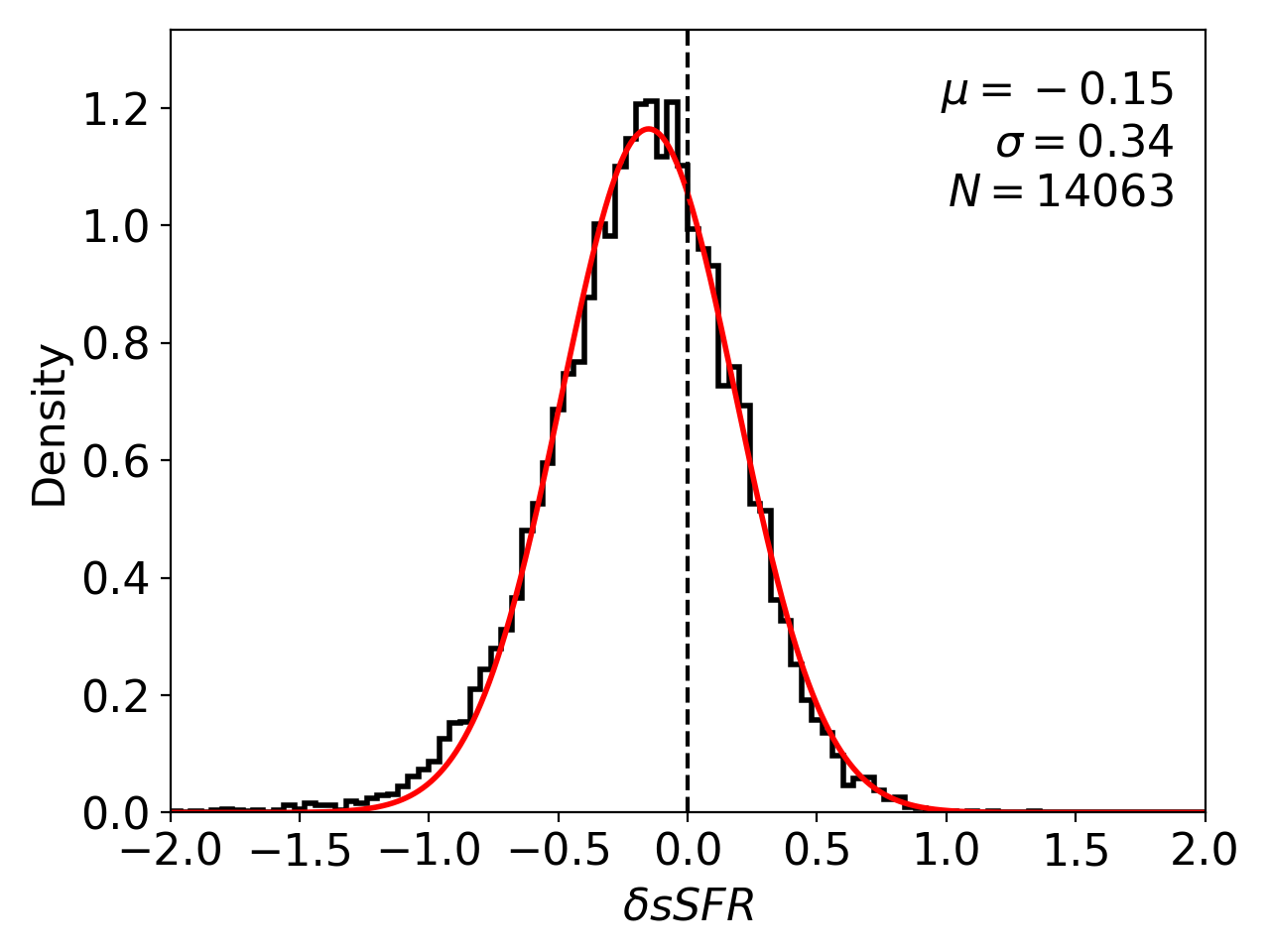}
        \caption{}
        \label{fig:ssfr_puertas}
\end{subfigure}\hfill
\caption{Comparison between SFR (a), stellar mass (b), and sSFR (c) from this study and \cite{puertas2017aperture} for the $N$ common sources. The red lines show the best Gaussian fits with median $\mu$ and dispersion $\sigma$.}
\end{figure*}

\begin{figure}[htbp!]
  \centering
  \includegraphics[width=0.8\linewidth]{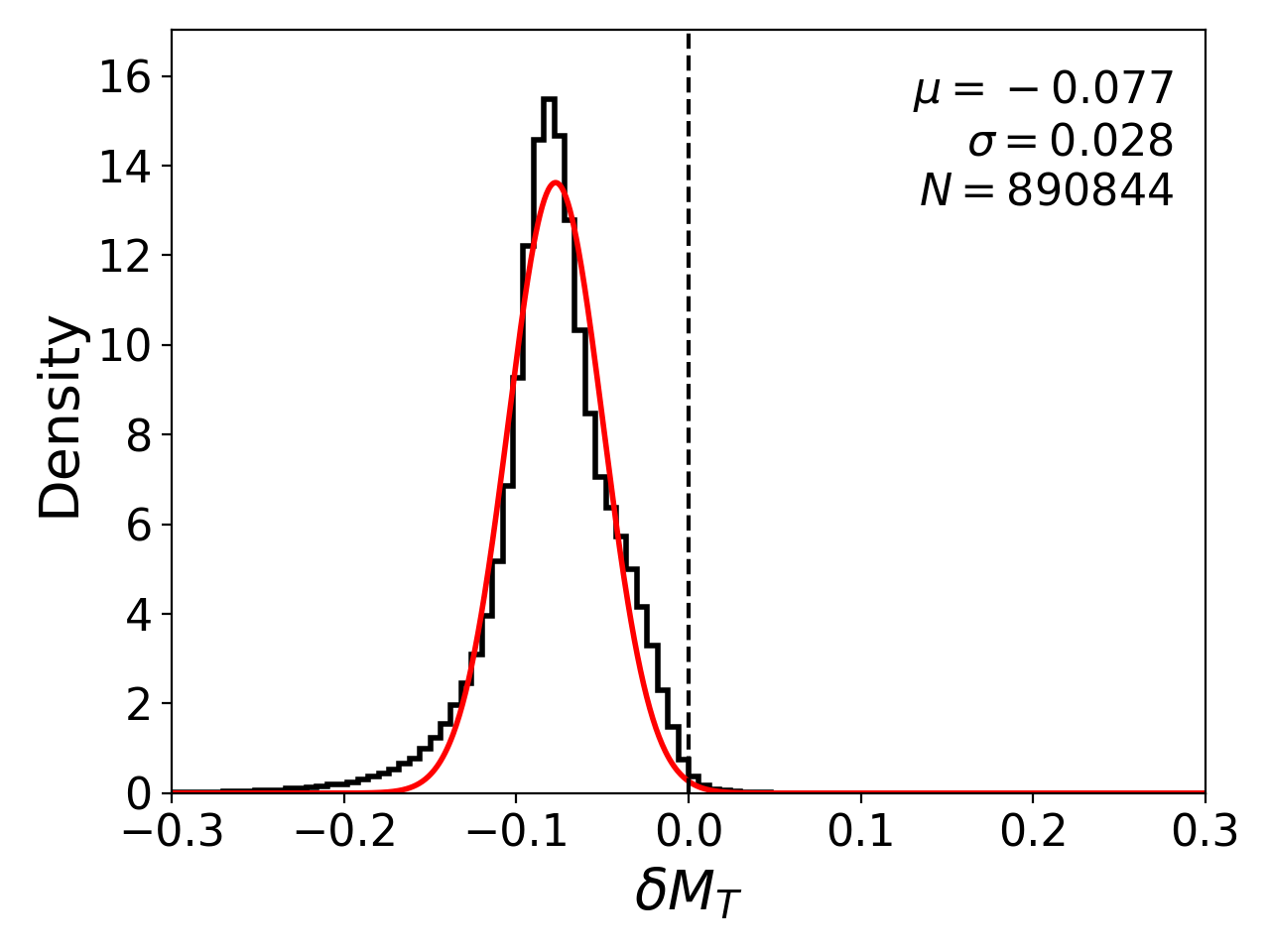}
\caption{Comparison of stellar masses from this study and \cite{taylor2011galaxy} for the $N$ common sources. The red line shows the best Gaussian fit with median $\mu$ and dispersion $\sigma$.}
\label{fig:clsj_taylor_mass}
\end{figure}

The stellar masses present a difference of $0.13$ dex and a dispersion of $0.2$ dex (Fig.~\ref{fig:mass_puertas}). The difference is compatible with typical systematical uncertainties due to the use of different models in stellar mass estimates ($\sim 0.3$ dex, \citealt{barro2011uv}). The stellar masses were further tested by applying the \cite{taylor2011galaxy} mass-to-light relation, $\log{(M_{*}}/{M_{\odot}}) = 1.15 + 0.70\,(g-i) - 0.4\,M_i$,
where $M_i$ is the absolute magnitude in the rest-frame $i$ band and $(g-i)$ is the rest-frame color. Both were derived from \texttt{CIGALE}. This relation can be used to estimate the stellar mass with a $0.1$ dex precision. We find an offset of $-0.08$ dex and a remarkable low dispersion of $0.03$ (Fig.~\ref{fig:clsj_taylor_mass}). The comparison with \citet{puertas2017aperture} includes the distance uncertainty from the photometric redshifts, while the comparison with \cite{taylor2011galaxy} only accounts for differences in the mass-to-light ratio inference.

The SFR and the sSFR are compared in Figs.~\ref{fig:sfr_puertas} and \ref{fig:ssfr_puertas}, respectively. The offsets obtained are $-0.01$ dex for SFR and $-0.15$ dex for sSFR, while the dispersions are $0.44$ dex and $0.34$ dex, respectively. The SFR is measured without bias, and the sSFR offset reflects the difference found in the stellar mass estimate. Because the distance uncertainty affects both the stellar mass and the SFR in the same way, the sSFR dispersion is lower.

The \citet{puertas2017aperture} sample is biased toward SF galaxies with $r \lesssim 18$ mag (Appendix~\ref{app:dp17}). We divided the sSFR distribution in Fig.~\ref{fig:ssfr_puertas} by the uncertainty derived from \texttt{CIGALE}. The resulting distribution is Gaussian, with $\sigma = 0.9$, close to the expected unity. This implies that \texttt{CIGALE} errors are reliable and overestimated by just $10$\%. The median of the sSFR uncertainties from \texttt{CIGALE} in the full J-PLUS sample increases from 0.3 dex at $15 < r < 16$ mag to 0.45 dex at $19 < r < 20$ mag. We conclude that the stellar mass and the sSFR can be properly estimated from J-PLUS optical photometry alone to within a factor of 2-3 at $r < 20$ mag.

\subsection{Quiescent and star-forming galaxies in J-PLUS DR3}
\begin{figure}[h]
    \centering        \includegraphics[width=0.475\textwidth]{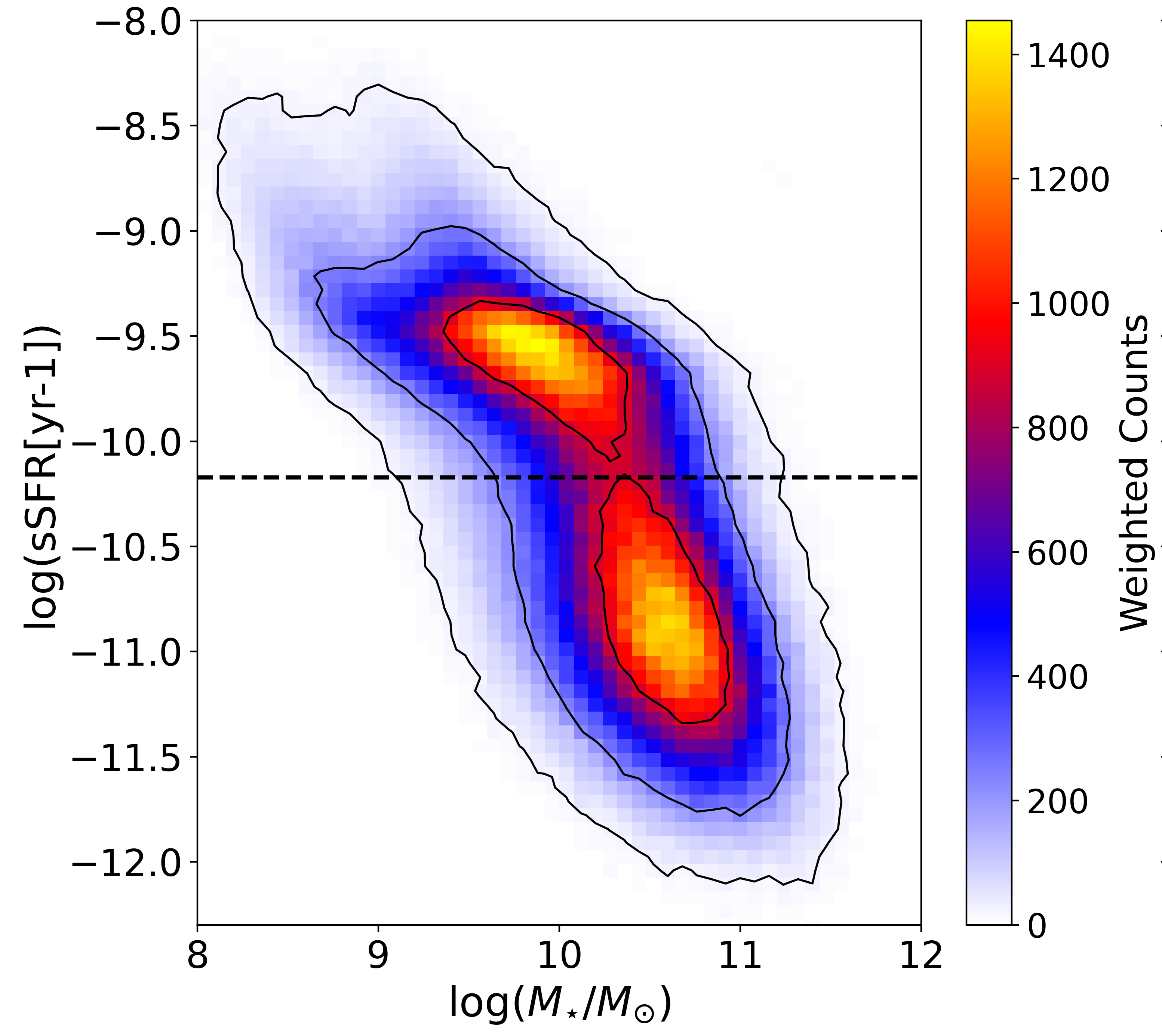}
        \caption{sSFR vs. stellar mass 2D \texttt{odds}-weighted histogram. The two maxima correspond to SF (higher sSFR values) and Q (lower sSFR values) populations. The limit between both populations is set to $\log\ {\rm sSFR} = -10.2$ dex (horizontal dashed line).}
        \label{fig:split}   
\end{figure}

The Q and the SF populations were selected using their position in the sSFR versus stellar mass plane. A 2D histogram of the $890$k galaxies under study, weighted by their \texttt{odds} parameter\footnote{The \texttt{odds} is defined as the integral of $P({\rm z})$ in the range $z_{\mathrm{phot}} \pm 0.03(1 + z_{\mathrm{phot}})$. The closer it is to one, the more concentrated is the photometric redshift estimate.}, was used to define the best selection (Fig.~\ref{fig:split}). Two populations are clearly visible on this diagram, corresponding to SF and Q galaxies. The former peaks at $\log\,M_{\star} = 9.9$ dex and $\log {\rm sSFR} = -9.5$ dex, the latter at $\log\,M_{\star} = 10.6$ dex and $\log {\rm sSFR} = -11.0$ dex. The limit of both populations was determined by identifying the relative minimum of the sSFR histogram with a double Gaussian distribution. We obtained $\log\,\rm{sSFR_{lim}} = -10.2$ dex, with no dependence on stellar mass. In the following, Q galaxies are defined as those with an sSFR lower than sSFR$_{\rm lim}$ and as SF otherwise. The initial sample was split into $488\,369$ SF galaxies and $402\,475$ Q galaxies. 

Other approaches are possible for splitting the global population into Q and SF galaxies. For example, one can use the rest-frame $(U-V)$ versus $(V-J)$ color-color diagram \citep{williams2009uvj} and their ultraviolet and infrared extensions \citep[e.g.,][]{leja2019beyond}. In addition, these diagrams can include dusty SF galaxies in the Q sample \citep{diaz2019stellar}. The analysis performed by \citet{leja2019beyond} suggests that the usual selection in the $(U-V)$ versus $(V-J)$ color-color diagram is equivalent to a $\log\,\rm{sSFR_{lim}} = -10.0 \ yr^{-1}$ selection, which is close to our inferred value.

The photometric and derived physical properties of the galaxy sample analyzed in this work were compiled into a catalog, a subset of which is shown in Table~\ref{tab:catalog_columns_vizier}. The catalog includes J-PLUS DR3 photometric measurements and stellar population parameters, such as stellar masses, SFRs, and rest-frame colors, estimated using \texttt{CIGALE}. In the next section, the methodology to determine the SMF for the total, SF, and Q populations is explained.

\section{Estimation of the stellar mass function}\label{sec:method}
Once we obtained good-quality stellar masses for Q and SF galaxies, we obtained the SMF of J-PLUS DR3. We followed the method described in \cite{pozzetti2010} for estimating mass completeness, using the $1/V_{max}$ technique and bootstrapping to estimate errors.

\subsection{Stellar mass completeness}\label{sec:completeness}
\begin{figure}
  \centering
  \includegraphics[width=\linewidth]{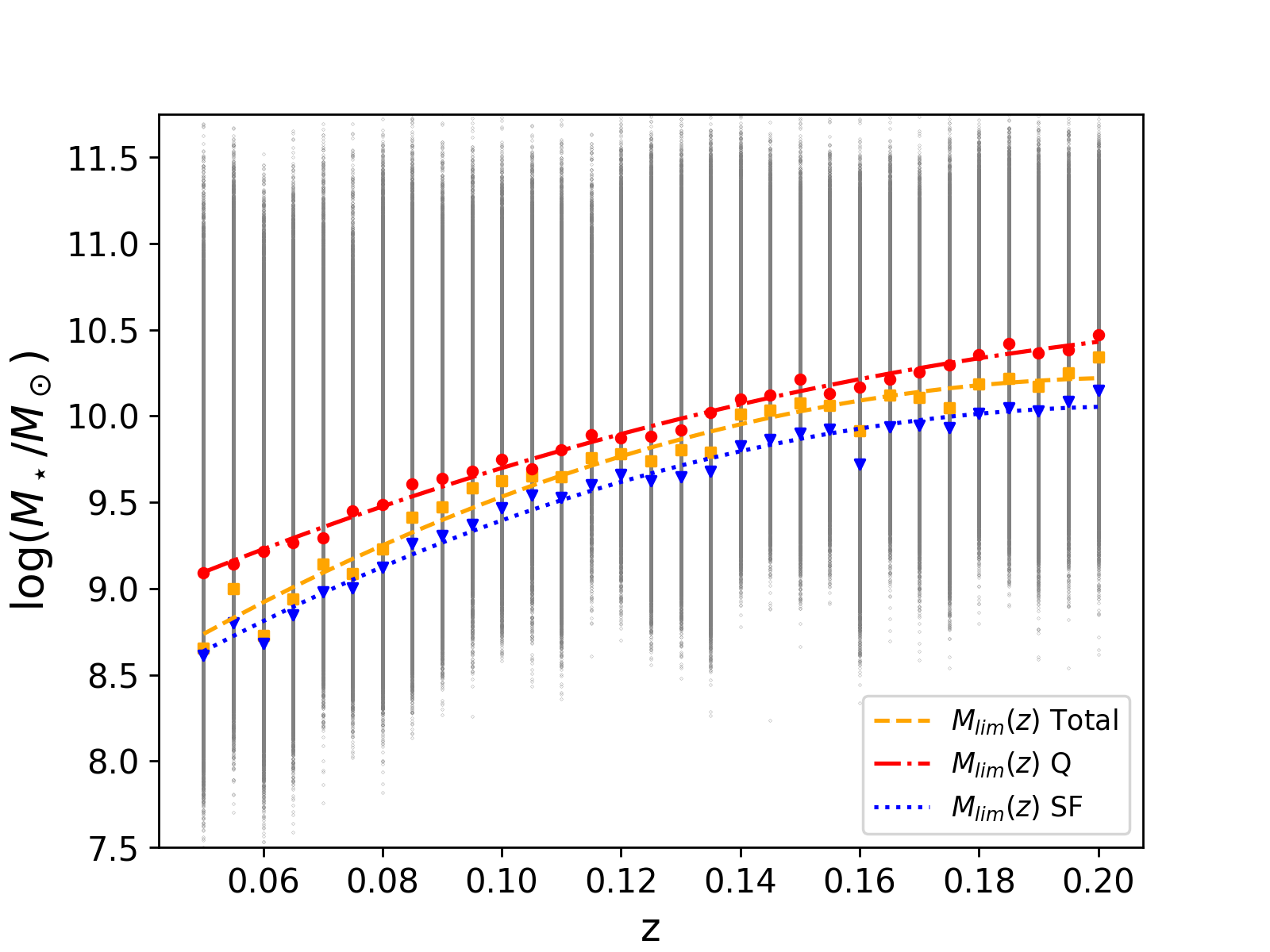}
  \caption{Stellar mass completeness limit $M_{\rm lim}(z)$ for the total, Q, and SF galaxy populations (orange, red, and blue dots, respectively). Second-order polynomial fits are shown for the total (dashed orange), Q (dash-dotted red), and SF (dotted blue) populations. 
}
  \label{fig:M_lim_2}
\end{figure}

We estimated the stellar mass completeness limit as a function of redshift, \( M_{\rm lim}(z) \), following the method introduced by \citet{pozzetti2010}. This approach computes, for each galaxy, the minimum stellar mass for an apparent magnitude equal to the survey’s limiting magnitude. The resulting function \( M_{\rm lim}(z) \) defines the stellar mass above which the sample is considered complete at each redshift. This correction is necessary in magnitude-limited surveys, which can include low-mass galaxies if they are sufficiently luminous, but can exclude massive galaxies that are too faint to be detected. As a result, the stellar mass completeness depends on both redshift and the distribution of mass-to-light ratios in the sample. Separate completeness limits were therefore derived for different galaxy populations (e.g., SF and Q). To determine the effective limiting magnitude of the J-PLUS DR3 sample, \( m_{\rm lim} \), we examined the distribution of the $r$ apparent magnitudes in the redshift bins. In each bin, we deselected the faintest 20\% of galaxies and retained the brightest among them. This yields a conservative estimate of the survey limit with \( m_{\rm lim} = 20.1 \), consistent with our magnitude cut.
Following \citet{pozzetti2010}, we computed for each galaxy $i$ a limiting stellar mass $M_{{\rm lim},i}$, defined as the mass it would have at its redshift if its apparent magnitude were equal to the survey limit $m_{\rm lim}$:
\begin{equation}
    \log M_{{\rm lim}, i} = \log M_i + 0.4 \times (m_i - m_{\rm lim}),
    \label{eq:M_lim_i}
\end{equation}
where \( M_i \) and \( m_i \) are the stellar mass and observed magnitude of the galaxy. This relation assumes that luminosity \( L \) is proportional to flux \( F \), and that \( M/L \) remains constant under rescaling. Next, we divided the sample into redshift bins of width $\Delta z = 0.005$,  matching the native redshift grid used by \texttt{LePhare} and smaller than the typical scatter of our photometric redshifts, $\sigma_{z}/(1+z) \simeq 0.011$. In each bin, we selected the faintest 20\% of galaxies in the $r$ magnitude and computed their individual \( M_{{\rm lim}, i} \) values using Eq.~(\ref{eq:M_lim_i}). From this subsample, we took the 95th percentile of the \( M_{{\rm lim}, i} \) distribution. This value represents the mass above which 95\% of the faintest galaxies would still be included in the sample and was used as the completeness threshold in that redshift bin. This process was repeated in all redshift bins, resulting in a set of completeness values that are fitted with a second-order polynomial. The fitted function \( M_{\rm lim}(z) \) defines the redshift-dependent stellar mass limit, which we computed separately for the total, SF, and Q galaxy populations. The results are shown in Fig.~\ref{fig:M_lim_2}.
After applying the completeness cut, we excluded all galaxies with \( M < M_{\rm lim}(z) \). For each remaining galaxy, its maximum redshift \( z_{{\rm max}, i} \) was computed as the redshift at which its stellar mass equals \( M_{\rm lim}(z) \). This was used in the estimation of volume-limited statistics such as \( V_{\max} \), following the procedure illustrated in Fig.~4 of \citet{weigel2016stellar}.

The final mass-complete sample consists of $302\,393$ SF and $376\,171$ Q galaxies. We note that the J-PLUS DR3 sample is complete at ${\rm z} \leq 0.2$ for stellar masses above $\log M_{\star}/M_{\odot} = 10.2$ dex, hereafter $\log M_{\star}$.

\subsection{The 1/V\textsubscript{max} method}\label{sec:vmax}
We used the $1/V_{\rm max}$ technique from Schmidt (1968) to correct for Malmquist bias by weighting each galaxy by the maximum volume over which it could be detected. We defined $V_{{\rm max}, i}$ as the maximum volume in which a galaxy $i$ at redshift $z_i$ with stellar mass $M_{\star,i}$ could be detected. The number density $\Phi$ in mass bin $j$ was computed by summing over all $N_j$ objects in that bin:
\begin{equation}
    \Phi_j = \dfrac{1}{d\log M_{\star}}\sum_i^{N_j} \dfrac{1}{V_{{\rm max},i}}\ \ \ [{\rm Mpc}^{-3}\, {\rm dex}^{-1}],
\end{equation}
where $d\log M_{\star} \approx (\log M_{\star,{\rm max}} - \log M_{\star,{\rm min}})/n_{\rm bin}$, $\log M_{\star,{\rm max}} = 11.5$ dex is the maximum considered mass, $\log M_{\star,{\rm min}} = 9.0$ dex is the minimum considered mass, and $n_{\rm bin} = 10$ is the number of bins. The comoving volume $V_{{\rm max}, i}$ in a flat universe is \citep{hogg1999distance}
\begin{equation}
    V_{{\rm max},i}=\dfrac{4\pi}{3}\dfrac{\Omega^s}{\Omega^{\rm sky}}\left[d_c({\rm z}_{{\rm max},i})^3 - d_c({\rm z}_{{\rm min},i})^3 \right],
\end{equation}
with $\Omega^s = 2\,881\ {\rm deg}^2$ being the effective surface area of J-PLUS DR3 after masking. Here, the surface area of the entire sky is $\Omega_{\rm sky} = 41\,253\ {\rm deg}^2$, $d_c\,(\rm z)$ is the comoving distance at redshift ${\rm z}$ for the assumed cosmology, ${\rm z}_{{\rm min},i} = 0.05$, and ${\rm z}_{{\rm max},i}$ is determined as
\begin{equation}
    {\rm z}_{{\rm max}, i} = min ({\rm z}_{\rm max}^{s}, {\rm z}^{\rm mass}_{{\rm max}, i}).
    \label{eq:zmax}
\end{equation}
We defined \( z_{\rm max}^{s} = 0.20 \) as the upper redshift limit of our analysis. For each galaxy \( i \), we computed its individual maximum redshift \( z^{\rm mass}_{{\rm max}, i} \), which corresponds to the redshift at which the galaxy would become undetected due to the stellar mass limit, \( M_{\rm lim}(z) \), as described in Sect.~\ref{sec:completeness}. This was obtained by inverting the \( M_{\rm lim}(z) \) relation to determine the highest redshift at which the galaxy stellar mass would still be above the limiting mass. This inversion is only meaningful within the mass and redshift ranges where \( M_{\rm lim}(z) \) is defined and monotonic. As a result, galaxies whose stellar masses remain above \( M_{\rm lim}(z) \) throughout the entire redshift range do not yield a finite solution from the inversion. For these cases, we assigned \( z^{\rm mass}_{{\rm max}, i} = z_{\rm max}^{s} = 0.20 \).
\subsection{Error budget in the SMFs}
\label{sec:smf_error_budget}

Random uncertainties. We estimated the uncertainties in the number density of SMF, $\Phi_j$, via bootstrap resampling. We generated 100 realizations by drawing, with replacement, a sample of the same size as the input sample. For each mass bin, $\Phi_j$ was taken as the median across realizations, and the $1\sigma$ uncertainty as the standard deviation of the $\Phi_j$ distribution.

Systematic uncertainties. To account for systematics between the Q and SF separation, we recomputed the SMFs using sSFR thresholds shifted by $\pm 0.1$~dex and took the systematic error in each bin as half of the absolute difference between the two perturbed cases. Additional systematic uncertainties related to stellar population modeling were not explicitly propagated in our SMFs, but are expected to be comparable to those discussed in earlier studies (e.g., \citealt{marchesini2009evolution}).

\begin{figure}[t]
    \centering
    \includegraphics[width=0.48\textwidth]{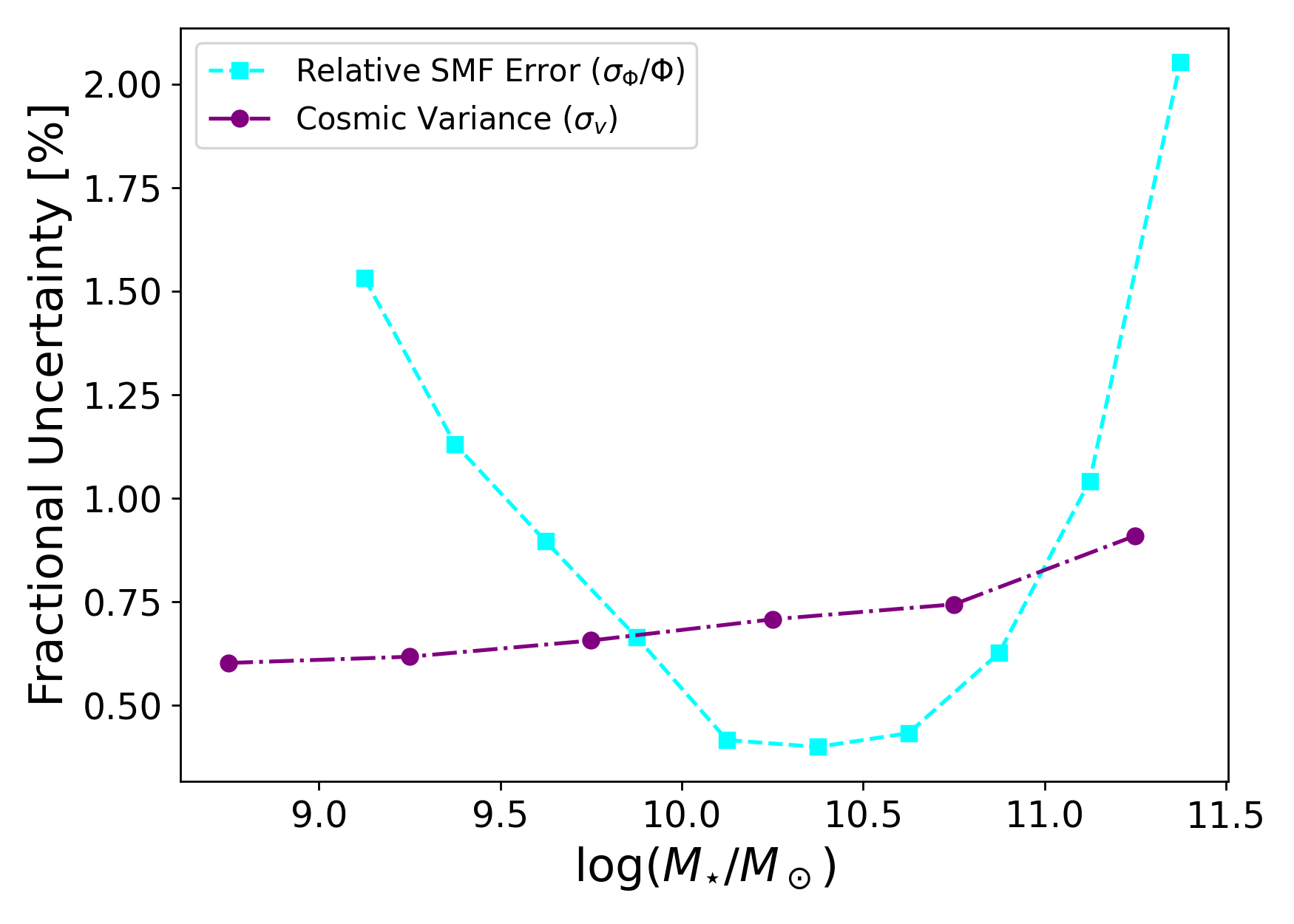}
    \caption{Cosmic variance for this study, estimated with the \citet{moster2011cosmic} prescription (purple circles), compared to the relative errors in the total SMF derived from bootstrapping (cyan squares).}
    \label{fig:cosmic_variance_smf}
\end{figure}

Cosmic variance. Cosmic variance ($\sigma_{\rm v}$) is a significant source of uncertainty in surveys with small sky coverage, such as pencil-beam fields ($\lesssim 1$~deg$^2$), where number-density fluctuations can be large owing to the limited cosmological volume probed \citep[e.g.,][]{diaz2024minijpas}. In contrast, J-PLUS DR3 covers an effective area of $2\,881$~deg$^2$, which should considerably mitigate the impact of cosmic variance on the measured SMF. We estimated $\sigma_{\rm v}$ using the framework of \citet{moster2011cosmic}, which models cosmic variance based on $\Lambda$CDM predictions and incorporates galaxy bias via a halo occupation distribution. From their Table~2, we adopted the root cosmic variance for dark matter, $\sigma_{\rm dm} = 0.181$, obtained for the COSMOS survey, and rescaled it to the J-PLUS DR3 area using their Equation~(12), correcting for the difference in survey area (from $1.96$~deg$^2$ to $2\,881$~deg$^2$). Figure~\ref{fig:cosmic_variance_smf} shows that $\sigma_{\rm v}$ increases with stellar mass, as expected since massive galaxies are more strongly clustered and typically reside in dense environments. We find that for $\log(M_\star/\mathrm{M}_\odot) < 11$, $\sigma_{\rm v}$ remains below $1$\%, reaching $\sim 1$\% only at $\log(M_\star/\mathrm{M}_\odot) \sim 11$. Over the $9.8 < \log(M_\star/\mathrm{M}_\odot) < 11.0$ range, cosmic variance is slightly larger than the statistical uncertainties obtained via bootstrapping; nevertheless, it remains small in absolute terms. Because cosmic variance arises from large-scale density fluctuations, it is highly correlated across stellar-mass bins and, to first order, acts as a global normalization uncertainty on the SMF (i.e., primarily on $\phi_\star$), rather than as independent random noise in each bin \citep{smith12cov,lopez17lf,diaz2024minijpas}. We therefore do not add $\sigma_{\rm v}$ in quadrature to the per-bin error bars. Rather, we treated it as a small, global systematic uncertainty on the SMF normalization.

To sum up, our error budget includes: (i) random uncertainties from finite sampling and the $1/V_{\max}$ weighting captured by the bootstrap resampling; (ii) a systematic component associated with the adopted separation between the Q and SF galaxies, estimated by varying the sSFR threshold by $\pm 0.1$~dex; and (iii) an additional, stellar-mass correlated contribution from cosmic variance, $\sigma_{\rm v}$, which primarily affects the overall normalization of the SMF. The final SMF measurements and their total statistical and systematic uncertainties are summarized in Appendix~\ref{app:smf}.

\subsection{Schechter function}
The observed SMFs can be parameterized with a \cite{schechter1976analytic} function, given by
\begin{equation}
    \Phi(M)\,dM = \Phi^{*} \left(\frac{M}{M^*}\right)^{\alpha} \exp\left(-\frac{M}{M^*}\right)\, dM,
    \label{eq:sch1}
\end{equation}
where $M^*$ is the stellar mass at which the function transitions from a simple power law with slope $\alpha$ at lower masses into an exponential function at higher masses. The normalization $\Phi^{*}$ corresponds to the number density at $M^*_\star$.\
In our case, for SMFs, it is better to work in $\log\,M_{\star}$ space and write the Schechter function as
\begin{equation}
    \Phi\,d\log M = \ln(10)\,\Phi^* \times e^{-10^{\log M - \log M^*_{\star}}} \times \left(10^{\log M - \log M^*_{\star}}\right)^{\alpha + 1} d\log M.
\end{equation}
We obtained the Schechter functions from the SMF input stellar masses and stellar mass errors from SED fitting and $1/V_{max}$ from the \cite{pozzetti2010} method. The \citet{obreschkow2018eddington} code fits the SMFs using the modified maximum-likelihood method, treating the Schechter function as a generative distribution and convolving it with the stellar-mass error distribution to correct for Eddington bias and selection effects. The likelihood was evaluated in $\log M_\star$ space using the stellar masses and $1/V_{\max}$ weights described in Sect.~\ref{sec:vmax}. Parameter uncertainties were estimated by bootstrap, resampling the galaxy catalog 100 times. Although a double-Schechter form is often adopted for the local SMF (e.g.,\citealt{baldry2012galaxy,wright2018gama}), it does not yield a statistically meaningful improvement over a single Schechter within our fitted mass range. The Bayesian evidence, evaluated over the mass-complete interval, consistently favors the single Schechter for the total, SF, and Q SMFs (Q-SMFs). We therefore adopted a single Schechter function throughout; for more details see Appendix~\ref{app:schechter_ic}.

\section{Results}
\subsection{Stellar mass function in J-PLUS DR3}
\begin{figure*}[htbp!]
\begin{subfigure}{.43\linewidth}
  \includegraphics[width=\linewidth]{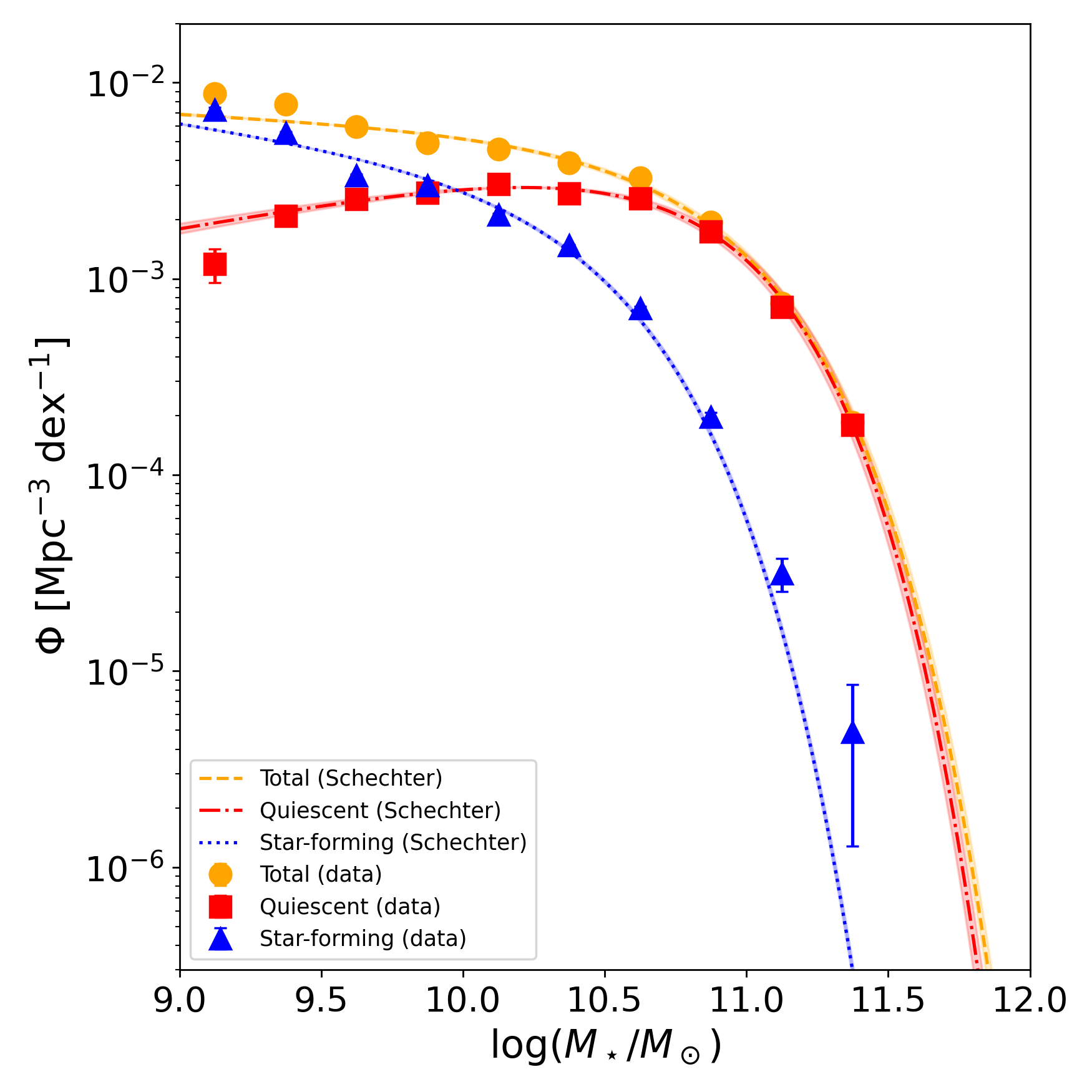}
  \caption{}
  \label{fig:schechter_jplus_smf}
\end{subfigure}\hfill
\begin{subfigure}{.43\linewidth}
  \centering
  \includegraphics[width=\linewidth]{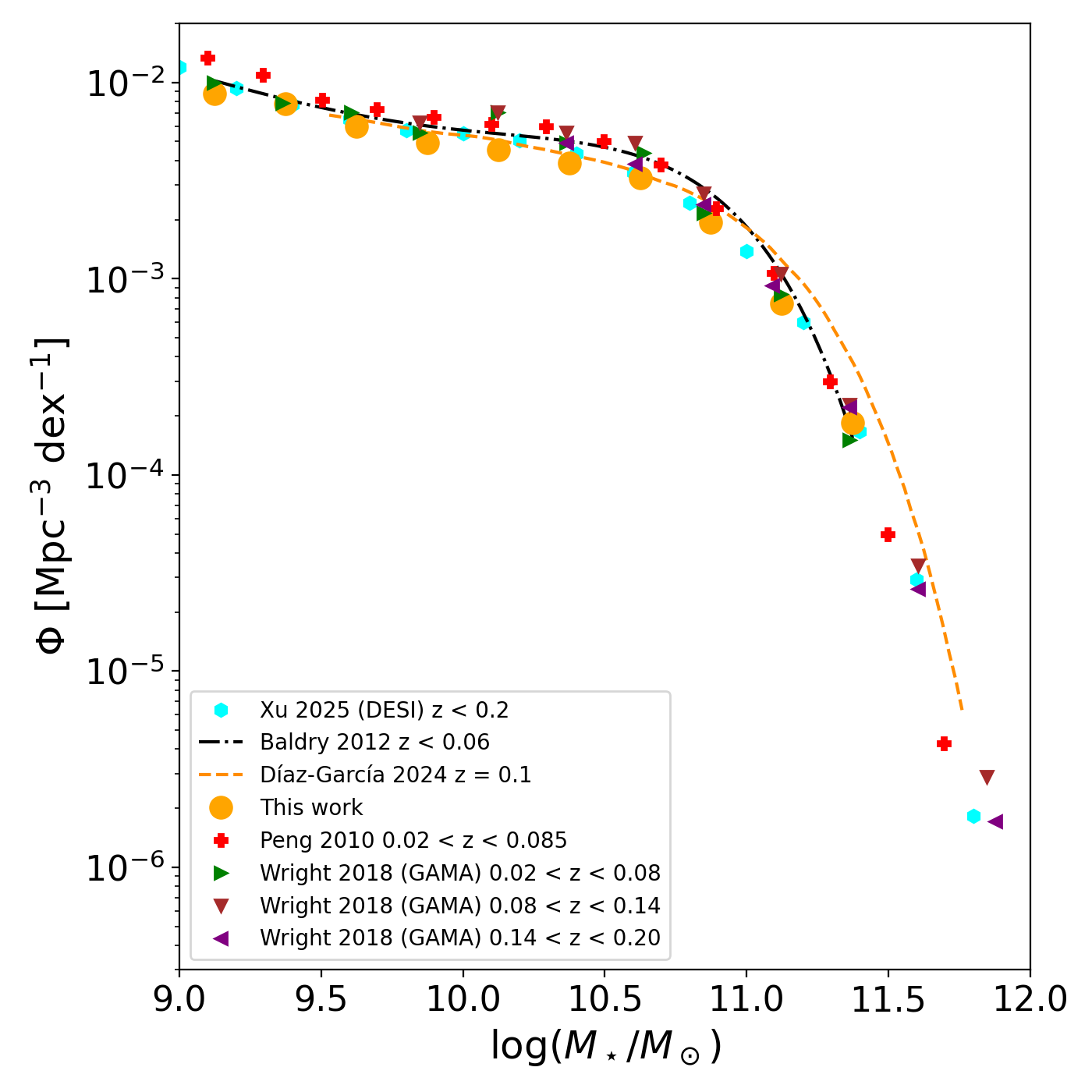}
  \caption{}
  \label{fig:total_smf}
\end{subfigure}\hfill
\begin{subfigure}{.43\linewidth}
  \includegraphics[width=\linewidth]{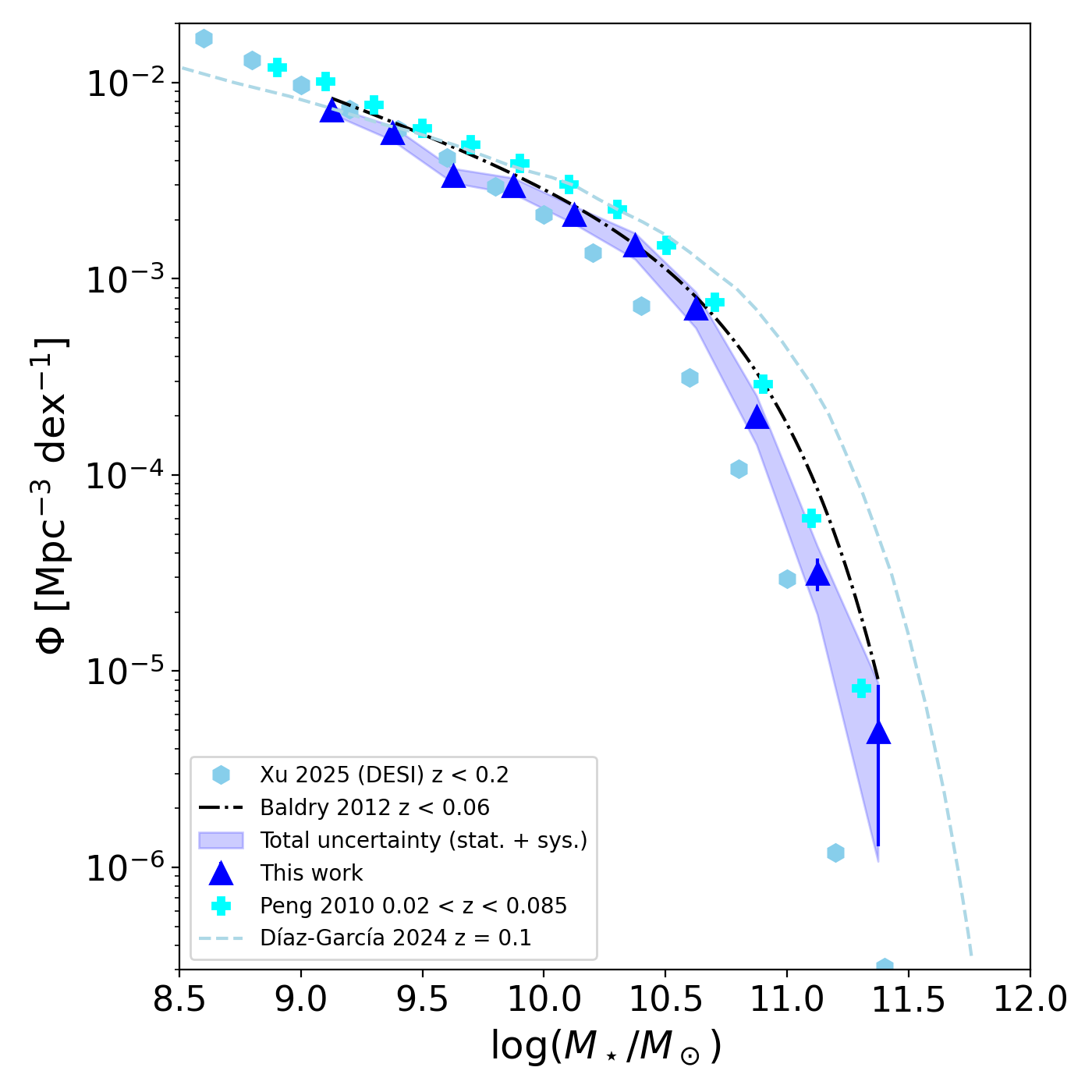}
  \caption{}
  \label{fig:sf_smf}
\end{subfigure}\hfill
\begin{subfigure}{.43\linewidth}
  \includegraphics[width=\linewidth]{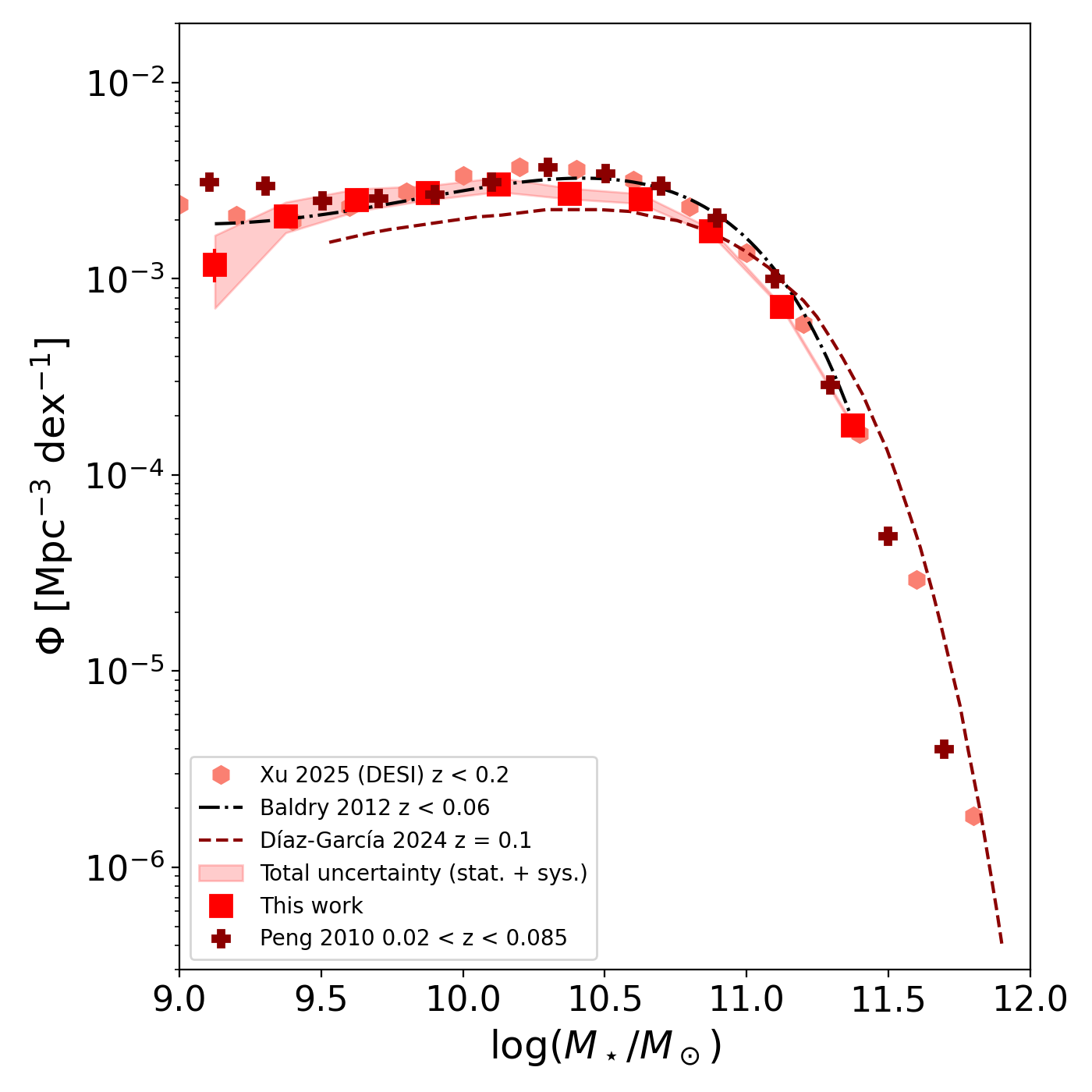}
  \caption{}
  \label{fig:q_smf}
\end{subfigure}\hfill
\caption{(a) Total, Q, and SF-SMFs from this study shown as orange circles, red squares, and blue triangles, respectively, with bootstrapping errors. Schechter fits are shown as dashed orange, dash-dotted red, and dotted blue lines. 
(b) Total SMF from J-PLUS DR3 (orange circles), compared to \citet{peng2010mass} (gray pluses), \citet{xu2025pac} (cyan hexagons), \citet{baldry2012galaxy} (purple points), and \citet{diaz2024minijpas} (dashed orange curve). GAMA SMFs from \citet{wright2018gama} are shown as green, brown, and purple triangles with different orientations. 
(c) SF-SMF from J-PLUS DR3 (blue triangles), with the blue-shaded region showing total uncertainty (statistical plus systematic). This is compared to \citet{baldry2012galaxy} (dash-dotted black curve), \citet{peng2010mass} (cyan pluses), \citet{xu2025pac} (light-blue hexagons), and \citet{diaz2024minijpas} (dashed light-blue curve).
(d) Q-SMF from J-PLUS DR3 (red squares), with the total uncertainty shown as a red-shaded region. This is compared to \citet{baldry2012galaxy} (dash-dotted black curve), \citet{peng2010mass} (dark red pluses), \citet{xu2025pac} (light-red hexagons), and \citet{diaz2024minijpas} (dashed dark-red curve).
}
\label{fig:smf_bins}
\end{figure*}

The measured SMF for the total, SF, and Q populations in J-PLUS DR3 at $0.05 \leq {\rm z}_{\rm phot} \leq 0.2$ are presented in Fig.~\ref{fig:schechter_jplus_smf}. We find that Q galaxies dominate at $\log M_{\star} > 10$ dex, with a larger number density of SF galaxies at lower stellar masses.
The best-fit parameters of the Schechter function are presented in Table~\ref{tab:bootstrap}. A single Schechter function provides a proper description of the data in the stellar mass range sampled, as shown in Fig.~\ref{fig:schechter_jplus_smf}. We find that the characteristic mass is higher by $0.4$ dex for Q galaxies, $\log M_{\star,{\rm Q}}^{*} = 10.80$ dex, than for the SF population, $\log M_{\star,{\rm SF}}^{*} = 10.43$ dex. This is accompanied by a higher characteristic density for Q galaxies. The faint-end slope is steeper for SF galaxies ($\alpha_{\rm SF} = -1.2$) than for Q galaxies ($\alpha_{\rm Q} = -0.7$). It is important to note that we also corrected the Eddington bias in the Schechter curve for the high masses of the SF-SMF, where our derived SF-SMF value lies slightly toward higher stellar masses. We observe small errors for all parameters and all populations, with values typically around $0.01$. Given the small uncertainties on $M_\star$ ($\sim 0.01$--$0.02$ dex), the $\sim 0.4$ dex difference between the characteristic masses of Q and SF galaxies is significant and confirms that the two populations occupy distinct regions of the SMF. The difference in faint-end slope ($\Delta\alpha \approx 0.5$) is also significant, supporting the picture in which low-mass galaxies are predominantly star forming, while the Q population is increasingly dominated by massive galaxies.
Our results imply that Q galaxies account for $45$\% of the number density in the local Universe at $\log M_{\star} > 9$ dex, but $75$\% of the stellar mass density.
A direct comparison between Schechter function fits and SMF must account for the strong dependence of the fit on the stellar mass range. Specifically, the faint-end slope, $\alpha$, cannot be reliably determined in the absence of low-mass galaxies. Although we also fixed $\alpha$ with a modified code based on \citet{obreschkow2018eddington} to constrain the remaining parameters, the resulting parameters that yield the Schechter function did not satisfactorily match the SMF. For this reason, we ultimately allowed all three Schechter parameters to vary freely.

\begin{table}[t!]
\caption{Bootstrapped Schechter parameters.}
\label{tab:bootstrap}
\centering
\renewcommand{\arraystretch}{1.3}
\begin{tabular}{lccc}
\hline\hline
Sample & $\log(\Phi_{*})$ & $\log(M^{*}/M_{\odot})$ & $\alpha$ \\
       & [dex$^{-1}$\,Mpc$^{-3}$] & & \\
\hline
Q     & $-2.632^{+0.012}_{-0.011}$ & $10.799^{+0.013}_{-0.013}$ & $-0.739^{+0.019}_{-0.019}$ \\
SF    & $-2.848^{+0.004}_{-0.004}$ & $10.427^{+0.004}_{-0.003}$ & $-1.204^{+0.005}_{-0.004}$ \\
Total & $-2.656^{+0.008}_{-0.012}$ & $10.872^{+0.011}_{-0.007}$ & $-1.073^{+0.007}_{-0.010}$ \\
\hline
\end{tabular}
\tablefoot{Uncertainties were derived from $10^{2}$ bootstrap realizations.}
\end{table}

\subsection{Fraction of the quiescent galaxies in J-PLUS DR3}
\begin{figure}[t]
    \centering
        \includegraphics[width=0.44\textwidth]{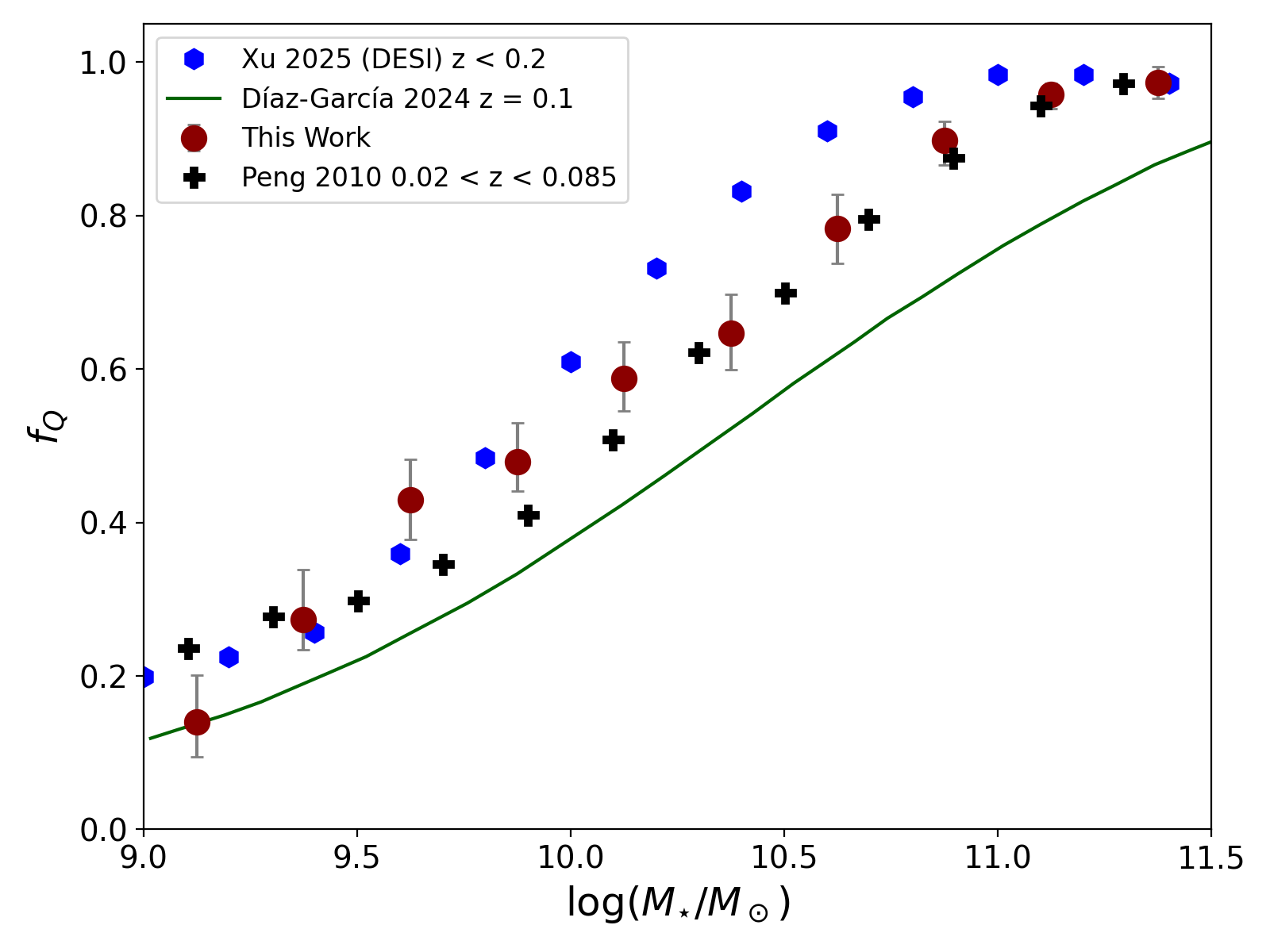}
        \caption{Q fraction as a function of stellar mass measured from different surveys. The red dots represent this work. The blue points are from DESI \cite{xu2025pac}, the black crosses are from \cite{peng2010mass}, and the green curve is from \cite{diaz2024minijpas}.}
        \label{fig:qfrac}
   
\end{figure}

The Q fraction, $f_{\rm Q}$, was computed from the binned SMFs as
\begin{equation}
    f_{\rm Q} = \dfrac{\Phi_{\rm Q}}{\Phi_{\rm Q} + \Phi_{\rm SF}}.
    \label{eq:qfrac}
\end{equation}
The Q fraction in J-PLUS DR3 is presented in Fig.~\ref{fig:qfrac}. Note that $f_{\rm Q}$ directly increases with stellar mass. An increase of $40$\% in the Q fraction per dex is observed, reaching $f_{\rm Q} > 0.95$ at $\log M_{\star} > 11$ dex. This tells us that low-stellar-mass galaxies are dominated by SF galaxies while high-stellar-mass galaxies are dominated by Q ones. This is a clear indication of the existence of an important mass-related quenching mechanism. The errors in the red fraction as a function of stellar mass were computed from the uncertainties in $\Phi_{\rm Q}$ and $\Phi_{\rm SF}$, and then propagated using Eq.~(\ref{eq:qfrac}). At the high-mass end, the small number of galaxies and the different mass-error distributions of Q and SF populations mean that residual Eddington bias and classification systematics might not be fully captured by our bootstrap errors. As discussed by \citet{weaver2023cosmos2020}, uncertainties in $f_{\rm Q}$ can be underestimated in this regime.

\section{Discussion}\label{sec:Discussion}
In this section, we interpret our SMF measurements from J-PLUS DR3 in the context of previous observational studies and theoretical models. We begin by comparing our total, Q, and SF-SMFs with literature results to assess consistency and systematic trends. We then examine the Q fraction as a tracer of galaxy evolution, evaluate the impact of cosmic variance, and compare our findings with predictions from the GAEA semi-analytic model (SAM).
\subsection{Stellar mass function}
In Fig.~\ref{fig:total_smf}, we compare the total SMF from J-PLUS DR3 with several measurements from the literature. The total SMF from the photometric study of \cite{peng2010mass} corresponds to slightly higher space densities for all stellar mass bins by $\sim$0.1~dex. The GAMA survey results from \cite{wright2018gama}, shown for three redshift bins ($0.02<z<0.08$, $0.08<z<0.14$, and $0.14<z<0.20$), also estimate higher space densities with respect to our determination, with typical offsets between 0.05 and 0.15~dex, especially at the high-mass end. In contrast, the SMF from the DESI spectroscopic survey \cite{xu2025pac} is in excellent agreement with our measurement, showing differences smaller than 0.05~dex over the full range. Finally, the total SMF from miniJPAS \citep{minijpas,diaz2024minijpas} is very similar to our total SMF, but at $\log(M_\star/\mathrm{M}_\odot) > 11$ it shows an excess of galaxies, reaching differences of up to $\sim$0.5~dex at the highest stellar masses. In Fig.~\ref{fig:sf_smf}, we compare the star-forming stellar mass function (SF-SMF) from J-PLUS DR3 with previous measurements from \citet{peng2010mass}, \citet{xu2025pac}, \citet{baldry2012galaxy}, and \citet{diaz2024minijpas}. Our SF-SMF closely follows the shape and normalization of the SMF from \citet{peng2010mass}, with a small offset of approximately 0.1~dex. Compared to the DESI measurement from \cite{xu2025pac}, our SF-SMF lies slightly higher, particularly at $\log(M_\star/\mathrm{M}_\odot) > 10.5$, where the difference can reach up to $\sim$0.15~dex. At lower stellar masses, both SMFs agree within the uncertainties. The SF-SMF from \citet{diaz2024minijpas} also lies above ours, similarly to \citet{peng2010mass} at $\log(M_\star/\mathrm{M}_\odot) < 10.5$; at higher stellar masses, it predicts a larger number of galaxies. In Fig.~\ref{fig:q_smf}, we compare the Q-SMF with results from \cite{peng2010mass} and \cite{xu2025pac}. Both literature SMFs lie slightly above our Q-SMF by up to 0.1~dex, but follow very similar trends. The agreement is particularly good with the DESI Q-SMF across most of the mass range. The Q-SMF from \cite{diaz2024minijpas} is below ours, by 0.08 dex, and the rest of the literature for $\log(M_\star/\mathrm{M}_\odot) < 11$ by 0.1 dex.

It is important to emphasize that once total uncertainties are considered, including statistical errors and systematics arising from the classification of Q and SF galaxies, our measurements remain fully consistent with previous studies. The shaded regions in Figs.~\ref{fig:sf_smf} and \ref{fig:q_smf} represent the total uncertainty envelopes, combining both components. These regions are similar in size to the scatter observed across the literature, especially at the low- and high-mass ends. This confirms the robustness of our methodology and supports the reliability of our population classification.

We conclude that the J-PLUS DR3 SMFs are in agreement with previous studies. The offsets between J-PLUS DR3 and previous SMFs (typically $\lesssim 0.1$--0.15 dex) are comparable to the scatter among literature measurements and can be explained by differences in survey selection, redshift range, and stellar population modeling. Given its wide effective area (2\,881 deg$^2$), intermediate depth, and 12-band optical photometry with seven narrowband filters, J-PLUS provides a particularly robust low-redshift SMF that we use as a reference for comparison with theoretical models and future surveys.
\subsection{Quiescent fraction}
In Fig.~\ref{fig:qfrac}, we present the Q fraction ($f_{\rm Q}$) as a function of stellar mass and compare it with two photometric studies: \citep{peng2010mass,diaz2024minijpas} and the spectroscopic DESI results from \cite{xu2025pac}. The uncertainties in our Q fraction include both the statistical errors from bootstrapping the SMFs of Q and SF galaxies at \(\log({\rm sSFR}) = -10.2\), and the systematic variations obtained by shifting this threshold by \(\pm0.1\) dex. Our measurement lies between \citep{peng2010mass} and \cite{xu2025pac} curves and most closely follows the shape of the photometric result. Compared to \cite{diaz2024minijpas}, our ($f_{\rm Q}$) is higher than theirs for $z = 0.1$.
A notable difference arises in the DESI quenched fraction, which rises more steeply from $\log(M_\star/\mathrm{M}_\odot)\sim9.5$ and reaches a plateau by $\log(M_\star/\mathrm{M}_\odot)\sim10.5$. In contrast, J-PLUS and \citet{peng2010mass} show a more gradual transition extending to $\log(M_\star/\mathrm{M}\odot)\sim11.0$. We quantified the agreement via point-by-point residuals after interpolating each curve to the J-PLUS mass grid, $\Delta f_{\rm Q}(M_\star)\equiv f_{\rm Q}^{\rm J\text{-}PLUS}-f_{\rm Q}^{\rm lit}$. For \citet{xu2025pac} (DESI), we obtain ${\rm RMSE}=0.086$ and $\langle\Delta f_{\rm Q}\rangle=-0.055$; for \citet{peng2010mass}$, {\rm RMSE}=0.061$ and $\langle\Delta f_{\rm Q}\rangle=0.022$; and for \citet{diaz2024minijpas}$, {\rm RMSE}=0.141$ and $\langle\Delta f_{\rm Q}\rangle=0.131$.
Overall, the rising $f_{\rm Q}$ trend with stellar mass observed in J-PLUS DR3 agrees with previous work and, being independent of SMF normalization, offers a robust probe of mass-dependent galaxy quenching.

\subsection{Comparison with theoretical models: The GAEA simulation}

The GAlaxy Evolution and Assembly (GAEA) model is a SAM of galaxy formation and evolution developed to run on top of dark matter halo merger trees from cosmological $N$-body simulations. Specifically, GAEA uses the Millennium Simulation \citep{springel2005gaea}, a large-volume $\Lambda$CDM simulation that traces the hierarchical growth of dark matter halos. The model includes detailed treatments of baryonic processes such as gas cooling, star formation, metal enrichment, stellar and AGN feedback, and environmental quenching. A key strength of GAEA lies in its chemically self-consistent enrichment scheme, which tracks individual elements with time-delayed feedback from type Ia and core-collapse supernovae \citep{de2014elemental}, along with an updated treatment for stellar feedback \citep{hirschmann2016gaea}. Star formation is linked to the molecular hydrogen content of the interstellar medium through an H$_2$-based law \citep{xie2017gaea}. The most recent version, GAEA2023, introduces updated prescriptions for galaxy quenching and satellite evolution and has been calibrated to match observed SMFs and star formation histories across cosmic time \citep{delucia2024gaea,fontanot2017gaea}. GAEA thus provides a comprehensive theoretical framework for interpreting statistical galaxy properties in the context of $\Lambda$CDM cosmology.

\subsubsection{Quiescent and star-forming galaxies in GAEA}
\begin{figure}[t!]
    \centering
    \includegraphics[width=0.44\textwidth]{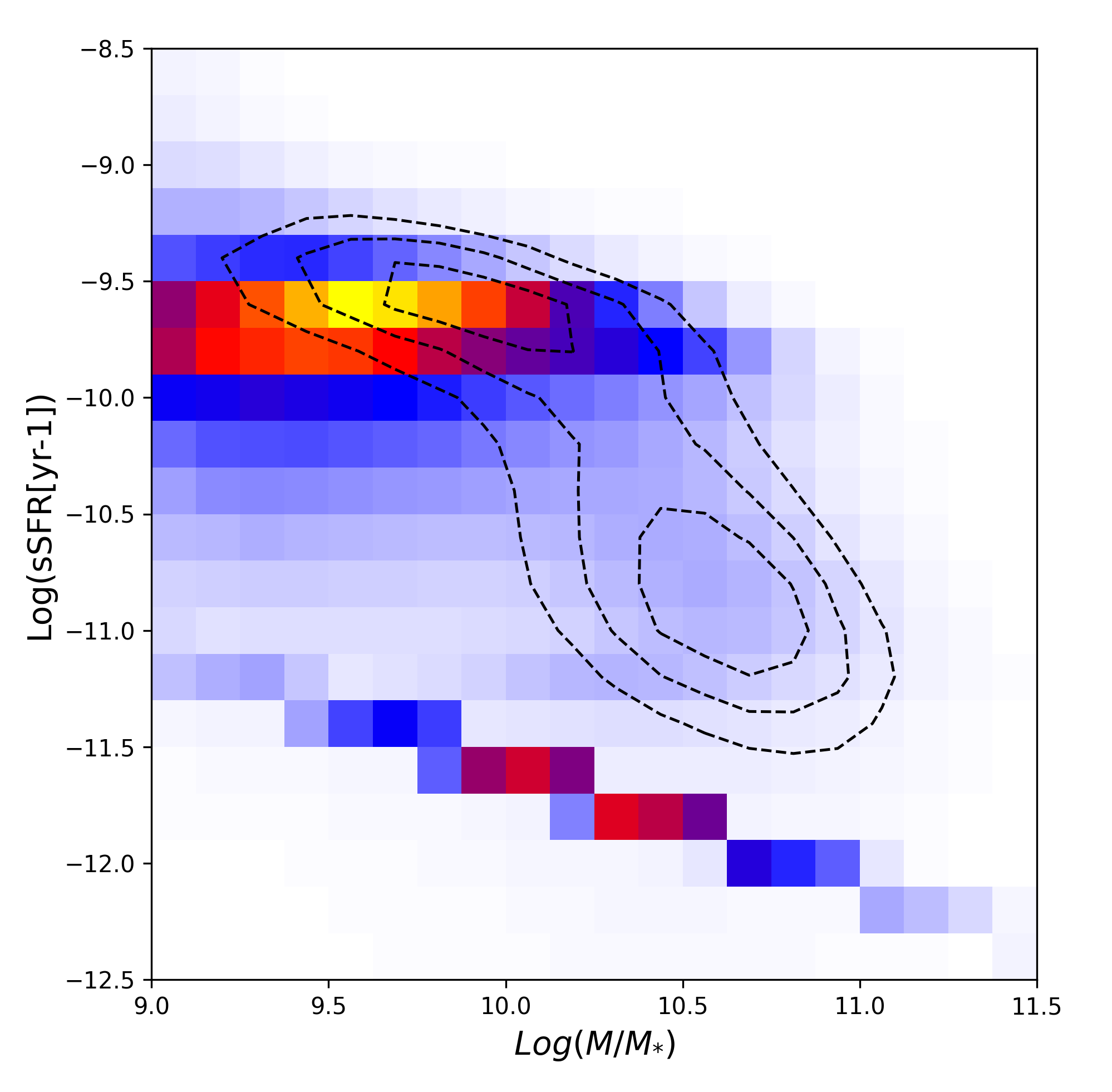}
    \caption{Specific SFR vs. stellar mass for the GAEA simulation, shown as a 2D histogram. The main sequence of SF galaxies is visible at $\log(\mathrm{sSFR}) > -11$, while Q galaxies with $\log(\mathrm{sSFR}) < -11$ form a diagonal stripe, reproducing the locus of Q galaxies in the SDSS (see \citealt{de2024tracing}). The dashed contours from our J-PLUS DR3 analysis are overplotted for comparison, as in Fig.~\ref{fig:split}.}
    \label{fig:split_gaea}
\end{figure}

To distinguish between Q and SF galaxies in the GAEA simulation, we adopted the sSFR as a classification criterion. GAEA associates an SFR estimate with each model galaxy, and in some cases these values can be very small. To take observational effects into account, we applied the same procedure as in \citet{de2024tracing} and assigned a new SFR value to each model galaxy with ${\rm SFR} < 10^{-4}$ using Eq.~\ref{eq:sfr_gaea}, which imposes an upper limit based on the empirical locus of Q galaxies in the SDSS:
\begin{equation}
    \log {\rm SFR} = 0.5 \times \log{M_\star} - 6.59,
    \label{eq:sfr_gaea}
\end{equation}
where $M_\star$ is in units of $M_\odot$. We applied this value to all GAEA galaxies with ${\rm SFR} < 10^{-4}$ and introduced a lognormal scatter of 0.25 dex to mimic observational uncertainties. In our J-PLUS DR3 data, the separation between Q and SF galaxies is defined by a threshold at $\log(\mathrm{sSFR}) = -10.2$. In contrast, GAEA studies often adopt a more conservative cut at $\log(\mathrm{sSFR}) = -11.0$ to define quenched systems. As shown in Fig.~\ref{fig:split_gaea}, both thresholds lie in the valley between the SF main sequence and the quenched population.

\subsubsection{Comparison of the stellar mass function}

\begin{figure*}[h!]
\centering
\begin{subfigure}{.33\linewidth}
  \includegraphics[width=\linewidth]{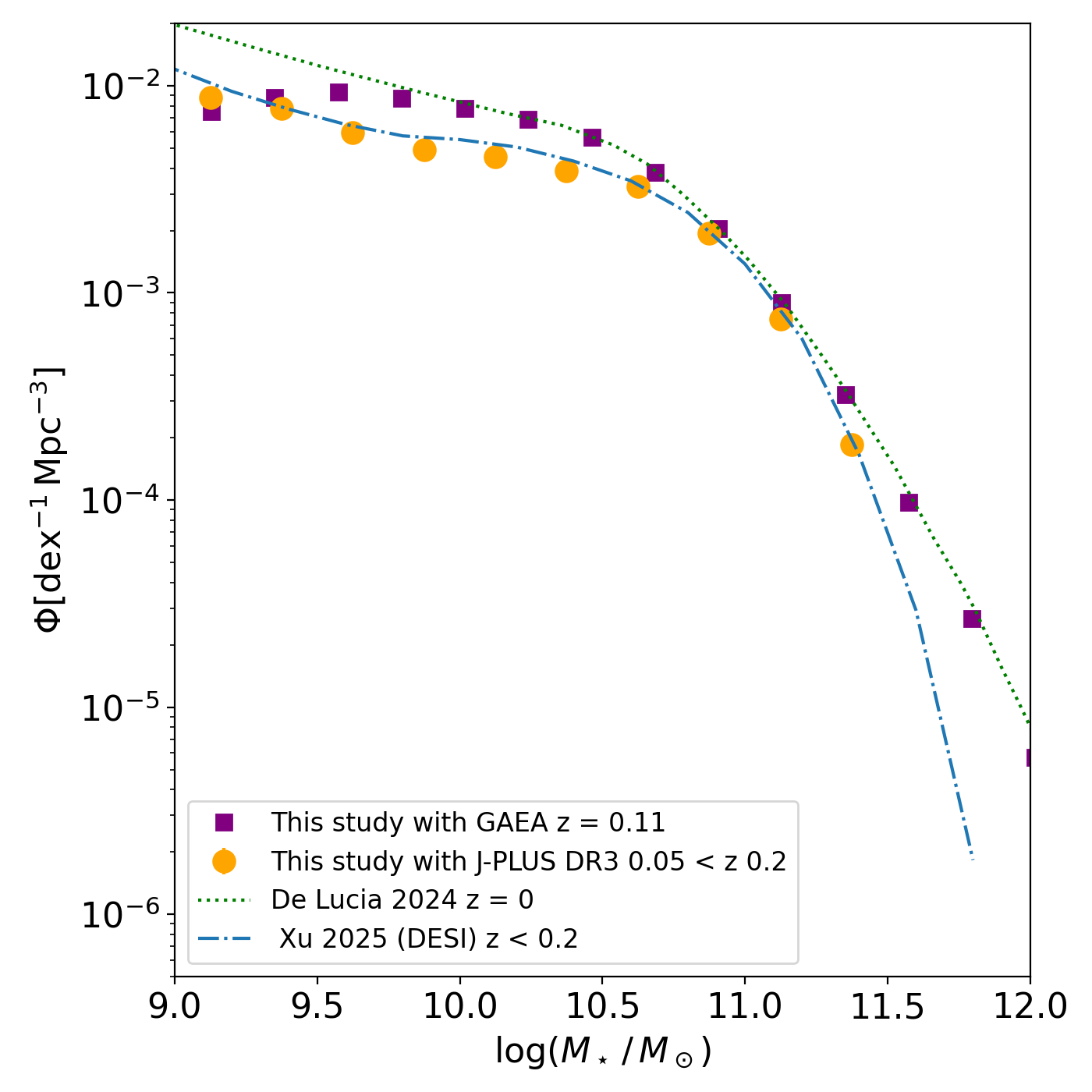}
  \caption{}
  \label{fig:total_smf_gaea}
\end{subfigure}
\begin{subfigure}{.33\linewidth}
  \includegraphics[width=\linewidth]{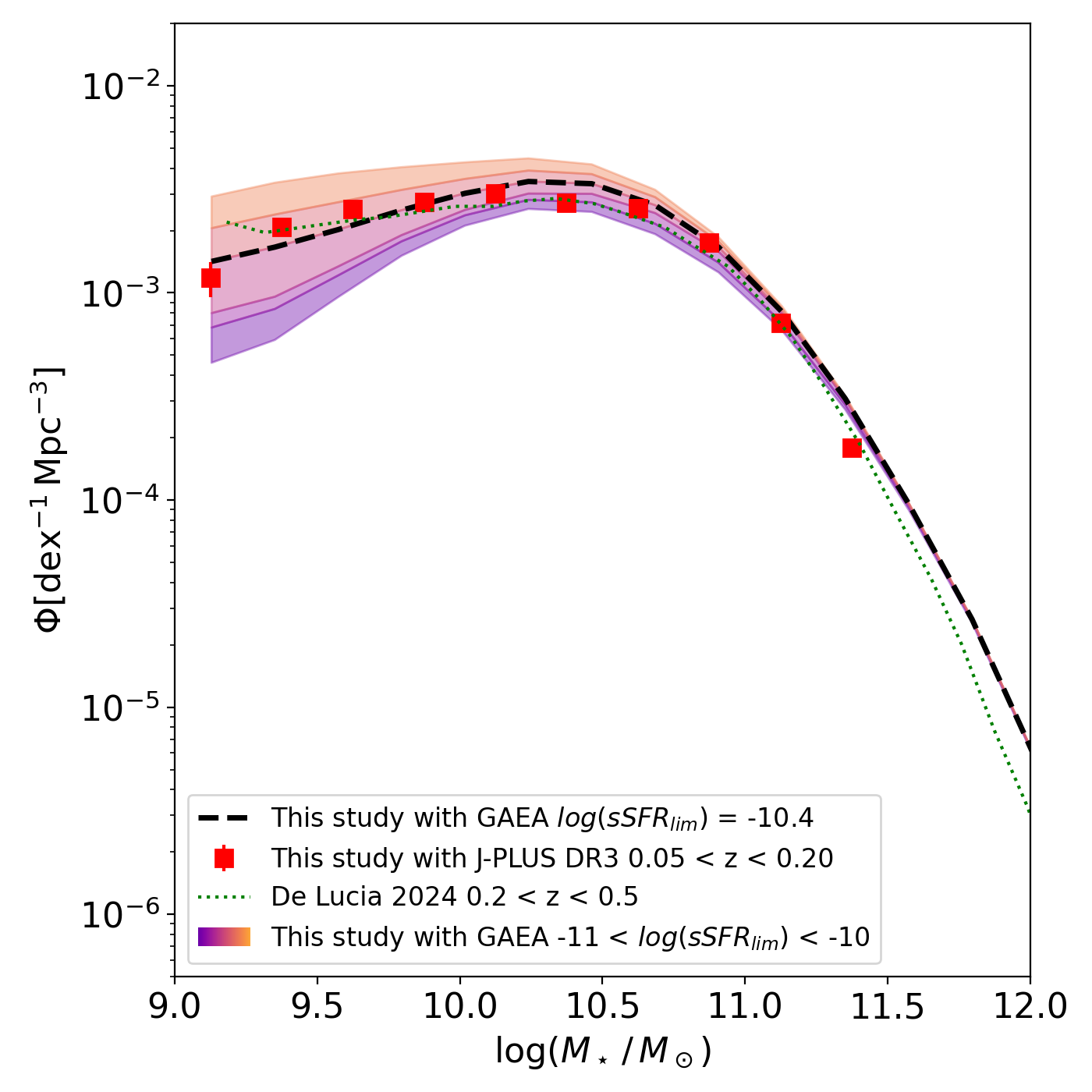}
  \caption{}
  \label{fig:q_smf_gaea}
\end{subfigure}
\begin{subfigure}{.33\linewidth}
  \includegraphics[width=\linewidth]{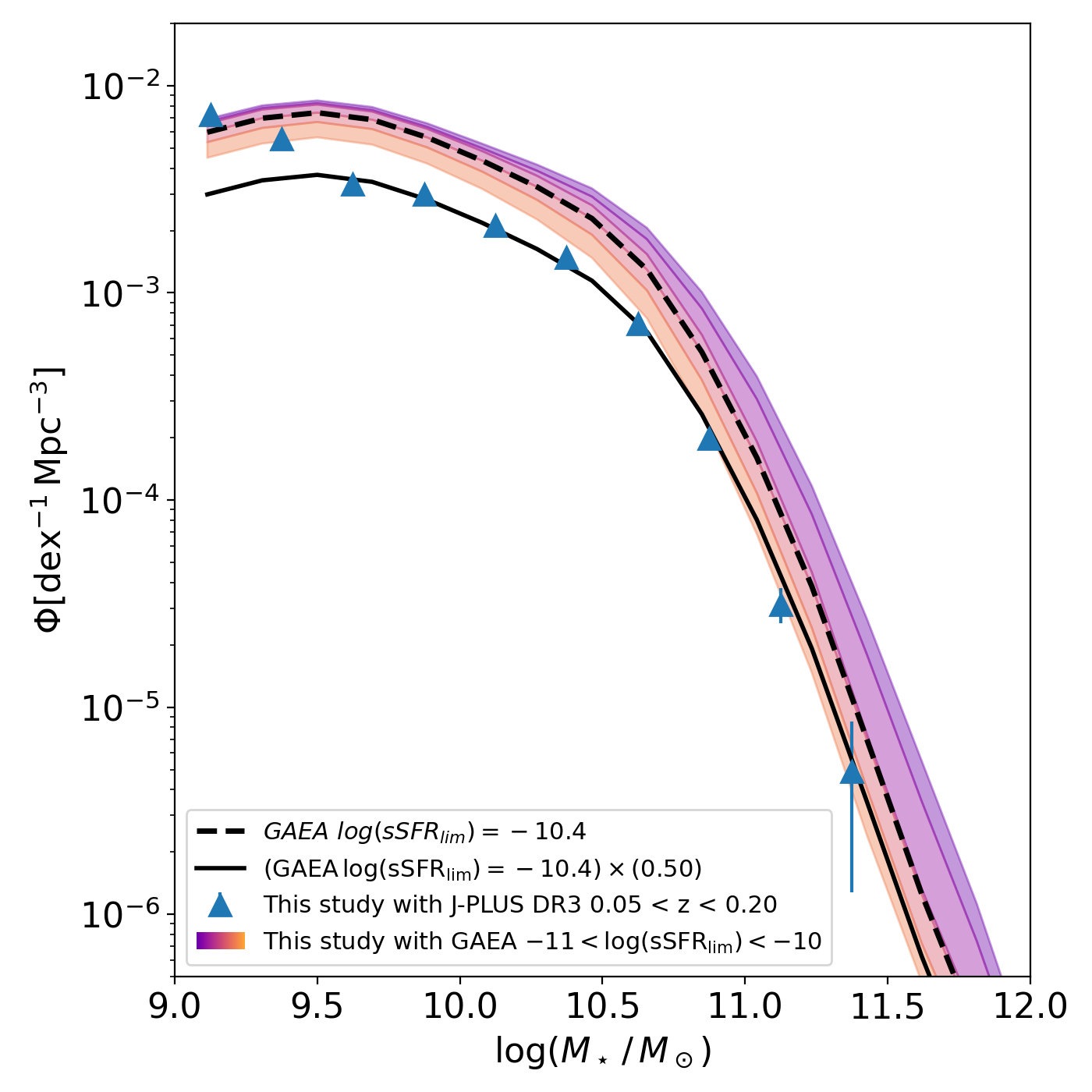}
  \caption{}
  \label{fig:sf_smf_gaea}
\end{subfigure}
\caption{Comparison of SMFs from J-PLUS DR3 and the GAEA simulation. (a) Total SMF from J-PLUS DR3 (orange circles), our GAEA-based sample using raw snapshot data and J-PLUS-like selection (purple squares), and the GAEA2023 model predictions published in \citet{delucia2024gaea} (dotted green curve). We also plot the total SMF from \cite{xu2025pac} (dash-dotted blue curve). (b) Q-SMFs derived using different thresholds in the range $-11 < \log({\rm sSFR}_{\rm lim}) < -10$ (shaded orange-purple curves), compared with J-PLUS DR3 (red squares) and GAEA2023 from \citet{delucia2024gaea} (dotted green curve). (c) SF-SMFs from J-PLUS DR3 (blue triangles), with GAEA predictions for varying $\log({\rm sSFR}_{\rm lim})$ (shaded orange-purple curves). The dashed black line corresponds to $\log({\rm sSFR}_{\rm lim}) = -10.4$, the value adopted for consistency with J-PLUS DR3, while the solid black line shows the same selection after applying a $-0.3$ dex stellar mass shift to the SF-SMF.}
\label{fig:smf_jplus_gaea}
\end{figure*}

We constructed SMFs from the GAEA simulation using the modeled stellar masses and sSFR-based population division. Galaxies were binned by stellar mass without applying observational corrections. We selected galaxies at $z = 0.11$ and $\log(M_\star/\mathrm{M}_\odot) > 9$. To mimic the J-PLUS selection, we applied an apparent magnitude limit of $r < 20$ mag and projected the simulation along the line of sight to account for observational geometry. Figure~\ref{fig:total_smf_gaea} presents the total SMF obtained from our GAEA sample, alongside the J-PLUS DR3 measurements and the GAEA2023 predictions from \citet{delucia2024gaea}. The two GAEA SMFs agree remarkably well across the intermediate- and high-stellar-mass ranges. The effect of the imposed magnitude limit on the predictions is also visible at low stellar masses. The reduced space densities reflect the loss of galaxies relative to the total GAEA SMF. At the low-mass end, our SMF falls below that of \citet{delucia2024gaea} due to the imposed apparent magnitude cut, which excludes faint galaxies and reduces completeness. Compared to J-PLUS DR3, the GAEA SMF lies systematically above the observations at $\log(M_\star/\mathrm{M}_\odot) > 10.0$, with an offset of up to $\sim$0.3~dex.

To investigate the division between Q and SF galaxies, we computed SMFs using six thresholds in the range $-11.0 < \log({\rm sSFR}_{\rm lim}) < -10.0$ dex. Figures~\ref{fig:q_smf_gaea} and \ref{fig:sf_smf_gaea} show the resulting Q and SF-SMFs compared with the J-PLUS DR3 measurements and the GAEA2023 SMFs from \citet{de2024tracing}. We find that Q-SMFs are particularly sensitive to the sSFR threshold at lower stellar masses, while the high-mass regime remains more stable. As expected, adopting a more negative sSFR limit increases the number of galaxies classified as SF, leading to a higher SF-SMF and a corresponding decrease in the Q-SMF. This effect is most pronounced below $\log(M_\star / \mathrm{M}_\odot) \sim 10.5$, where the population balance is more sensitive to classification. The best match to the J-PLUS DR3 SF-SMF occurs for thresholds near $\log({\rm sSFR}) = -10.0$, while the Q-SMF aligns better with $\log({\rm sSFR}) = -10.4$. The J-PLUS-derived separation value of $\log({\rm sSFR}) = -10.4$ dex thus represents a reasonable compromise for comparing populations in GAEA and J-PLUS. Nevertheless, the GAEA simulation predicts an excess of SF galaxies in the range $9.5 < \log(M_\star/\mathrm{M}_\odot) < 11.0$, which drives the total SMF above the observed values. Applying a stellar mass correction of $-0.3$ dex to the SF population in GAEA improves the agreement with J-PLUS DR3, as shown in Fig.~\ref{fig:sf_smf_gaea}. 

We find an offset between GAEA and J-PLUS DR3. While the SMF shapes agree well, the total SMF from GAEA2023 exceeds the observations by up to $\sim$0.3~dex at $\log(M_\star/\mathrm{M}_\odot) \sim 9.5$--$10.5$. Such differences are within the uncertainties discussed in \citet{delucia2024gaea}, who report comparable offsets when comparing GAEA2023 to the SMFs from \citet{muzzin2013evolution} and \citet{weaver2023cosmos2020}.
\subsubsection{Comparison of the quiescent fraction}

\begin{figure}[htb]
    \centering
    \includegraphics[width=0.45\textwidth]{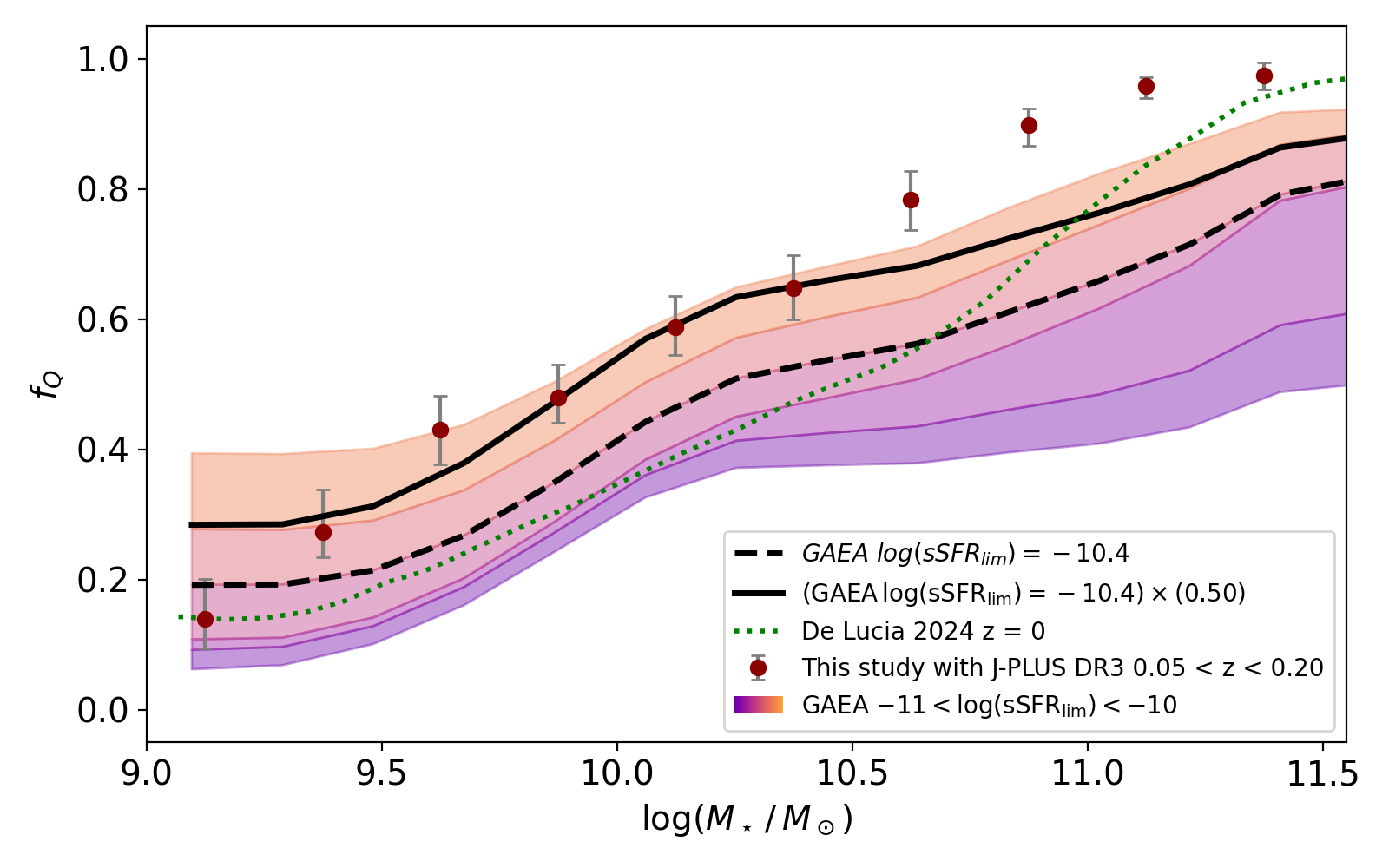}
    \caption{$f_Q$ from J-PLUS DR3 (red dots) compared to GAEA predictions for different $\log(\mathrm{sSFR}_{\mathrm{lim}})$ thresholds (color-coded lines). We also show the GAEA2023 results from \citet{delucia2024gaea} (dotted green line). The dashed black line shows the $f_Q$ for $\log(\mathrm{sSFR}_{\mathrm{lim}}) = -10.4$, and the solid black line applies a $-0.3$ dex stellar mass correction to the GAEA SF population.}
    \label{fig:qfrac_gradient}
\end{figure}

Figure~\ref{fig:qfrac_gradient} presents the Q fraction as a function of stellar mass, comparing results from J-PLUS DR3 with predictions from the GAEA simulation. We display GAEA $f_Q$ values for multiple sSFR thresholds, as well as the GAEA2023 $f_Q$ values from \citet{de2024tracing}, which remain consistently below the observed $f_Q$ over the full stellar mass range, reflecting the excess of SF galaxies in the simulation. This discrepancy is especially evident at intermediate and high masses. For the threshold $\log(\mathrm{sSFR}_{\mathrm{lim}}) = -10.4$ dex, consistent with the division adopted in our J-PLUS DR3 analysis, the GAEA predictions reproduce the observed Q fraction reasonably well at stellar masses below $\log(M_\star/\mathrm{M}_\odot) \sim 10.5$. At higher stellar masses, however, GAEA underpredicts the Q fraction. This indicates that a larger fraction of massive galaxies remain SF in the simulation. This discrepancy persists even after applying a $-0.3$ dex correction to the stellar masses of SF galaxies. Although the adopted threshold provides a reasonable basis for comparison, the results suggest that quenching is not fully efficient in GAEA at the high-mass end.

\section{Conclusions}\label{sec:Conclusions}
We presented the SMF of Q and SF galaxies at $0.05 \leq z \leq 0.2$ based on high-quality photometric data from J-PLUS DR3. Using a sample of $890\,844$ galaxies selected with $r_{0} \leq 20$ mag and $0.05 < z < 0.20$, we derived stellar masses and SFRs via SED fitting with \texttt{CIGALE} and separated galaxy populations based on their specific SFRs. The main conclusions of our study are as follows.

Stellar mass functions. We computed the SMFs for total, Q, and SF galaxies using the $1/V_{\mathrm{max}}$ method, accounting for stellar mass completeness and Eddington bias. All three SMFs are well described by single Schechter functions and are consistent with the literature. Quiescent galaxies dominate at $\log M_{\star} > 10$ dex, while SF galaxies are more numerous at lower stellar masses. Quiescent galaxies account for $45$\% of the number density in the local Universe at $\log M_{\star} > 9$ dex, but $75$\% of the stellar mass density.

Quiescent fraction. The fraction of Q galaxies increases by $40$\% per dex in stellar mass, as expected from mass-dependent quenching, reaching $f_{\rm Q} > 0.95$ at $\log M_{\star} > 11$ dex. Our measured Q fraction lies between the values obtained in previous studies, reinforcing the robustness of our classification.

Cosmic variance. Given the large effective area covered by J-PLUS DR3 ($2\,881\,\mathrm{deg}^2$), the impact of cosmic variance on our SMF measurements is minor. Using the prescription of Moster et al. (2011), we estimate relative uncertainties due to cosmic variance to be below 1\% for $\log M_\star < 11.5$ dex.

Comparison with simulations. We compared our observed SMFs with predictions from the GAEA SAM. GAEA2023 overpredicts the abundance of SF galaxies at $9.5 < \log M_\star < 11.0$ dex, leading to systematic excess in the total SMF and a corresponding deficit in the Q population. The offset reaches up to $\sim 0.3$ dex and persists despite matching sSFR thresholds and applying mass corrections. These differences are consistent with those reported in \citet{delucia2024gaea} and remain within the expected systematic uncertainties.

These results validate the scientific potential of J-PLUS DR3 for low-redshift galaxy evolution studies. The methods and findings presented here can be extended with J-PAS, which will provide deeper and higher-resolution photometry over a wider spectral range. The precision of the derived SMFs and Q fractions, combined with the large survey area, enables statistically robust analyses of environmental effects in future work.
\section*{Data availability}
The catalog associated with this article, described in Table~F.1, is available at the CDS via anonymous ftp to cdsarc.u-strasbg.fr (130.79.128.5) or via the CDS/VizieR online catalog service.
\begin{acknowledgements}
F.~D.~A.~B. acknowledge the funding from Erasmus IBERUS+ Student Mobility for Traineeships, and Programa Becas Ibercaja-CAI Estancias de Investigación for the research stay at INAF to work with G.d.L. and M.F. in the analysis of the GAEA simulations. The authors thank the interesting suggestions of the referee, which improved the paper.
J.~A.~F.~O., H.~D.~S., and A.~E. acknowledge the financial support from the Spanish Ministry of Science and Innovation and the European Union - NextGenerationEU through the Recovery and Resilience Facility (RRF) project ICTS-MRR-2021-03-CEFCA.
H.~D.~S. also acknowledges financial support by RyC2022-030469-I grant, funded by MCIU/AEI/10.13039/501100011033 and FSE+.
A.~L.~C. and P.~T.~R. acknowledge the financial support from the European Union - NextGenerationEU through the RRF program Planes Complementarios con las CCAA de Astrof\'{\i}sica y F\'{\i}sica de Altas Energ\'{\i}as - LA4.
J.~V.~M. acknowledges financial support by PID2022-136598NB-C32 grant.
L.~A.~D.~G. acknowledges financial support from the State Agency for Research of the Spanish MCIU through 'Center of Excellence Severo Ochoa' award to the Instituto de Astrofísica de Andalucía (CEX2021-001131-S) funded by MCIN/AEI/10.13039/501100011033 and to PID2022-141755NB-I00.
Based on observations made with the JAST80 telescope and T80Cam camera for the J-PLUS project at the Observatorio Astrof\'{\i}sico de Javalambre (OAJ), in Teruel, owned, managed, and operated by the Centro de Estudios de F\'{\i}sica del Cosmos de Arag\'on (CEFCA). We acknowledge the OAJ Data Processing and Archiving Unit (UPAD; \citealt{upad}) for reducing the OAJ data used in this work.Funding for the J-PLUS Project has been provided by the Governments of Spain and Arag\'on through the Fondo de Inversiones de Teruel; the Aragonese Government through the Research Groups E96, E103, E16\_17R, E16\_20R, and E16\_23R; the Spanish Ministry of Science and Innovation (MCIN/AEI/10.13039/501100011033 y FEDER, Una manera de hacer Europa) with grants PID2021-124918NB-C41, PID2021-124918NB-C42, PID2021-124918NA-C43, and PID2021-124918NB-C44; the Spanish Ministry of Science, Innovation and Universities (MCIU/AEI/FEDER, UE) with grants PGC2018-097585-B-C21 and PGC2018-097585-B-C22; the Spanish Ministry of Economy and Competitiveness (MINECO) under AYA2015-66211-C2-1-P, AYA2015-66211-C2-2, AYA2012-30789, and ICTS-2009-14; and European FEDER funding (FCDD10-4E-867, FCDD13-4E-2685). The Brazilian agencies FINEP, FAPESP, and the National Observatory of Brazil have also contributed to this Project.
\end{acknowledgements}
\bibliographystyle{aa}
\bibliography{aanda}

\appendix

\section{ADQL of the data}\label{app:adql}
The ADQL query used in this study is:
\lstset{breaklines=true, basicstyle=\ttfamily\small}
\begin{lstlisting}
SELECT 
  obj.TILE_ID,
  obj.number,
  obj.ALPHA_J2000 AS RA,
  obj.DELTA_J2000 AS DEC,
  obj.mask_flags,
  obj.MAG_AUTO AS MAG_AUTO,
  obj.MAG_ERR_AUTO,
  obj.class_star AS CLASS_STAR,
  lephare.photoz,
  lephare.odds AS ODDS,
  galclass.TILE_ID AS galclass_TILE_ID,
  galclass.number AS galclass_number,
  galclass.sglc_prob_star,
  mwe.TILE_ID AS mwe_TILE_ID,
  mwe.number AS mwe_number,
  mwe.ax AS ax,
  mwe.ax_err AS ax_err

FROM jplus.MagABDualObj AS obj

JOIN jplus.StarGalClass AS galclass
  ON obj.TILE_ID = galclass.TILE_ID
  AND obj.number = galclass.number

JOIN jplus.PhotoZLephare AS lephare
  ON obj.TILE_ID = lephare.TILE_ID
  AND obj.number = lephare.number

JOIN jplus.MWExtinction AS mwe
  ON obj.TILE_ID = mwe.TILE_ID
  AND obj.number = mwe.number

WHERE 
  lephare.photoz >= 0.05 
  AND lephare.photoz <= 0.2025
  AND obj.mask_flags[jplus::rSDSS] = 0
  AND galclass.sglc_prob_star < 0.5
  AND (obj.MAG_AUTO[jplus::rSDSS] - mwe.ax[jplus::rSDSS]) <= 20
\end{lstlisting}
\section{\texttt{CIGALE} configuration file}\label{app:cigale}
Below we summarize the modules and grid of parameters used in the \texttt{CIGALE} configuration file.
\lstset{breaklines=true, basicstyle=\ttfamily\small}
\begin{lstlisting}
#Configuration of the SED creation modules.
[sed_modules_params]
  
  [[sfhdelayed]]
  
    tau_main = 100., 1000., 5000.
    
    age_main = 3000., 5000., 7000., 10000., 13000.
    
    tau_burst = 5., 100., 500.

    age_burst = 5.,10.,50., 100., 500., 1000.

    f_burst =  0.001, 0.01, 0.2, 0.4

    sfr_A = 1.0

    normalise = True
 
  
  [[bc03]]
  
    imf = 1

    metallicity =  0.004, 0.008, 0.02

    separation_age = 10
       
  
  [[nebular]]

    logU = -2.0, -3.5

    zgas = 0.014, 0.022

    ne = 100

    f_esc = 0.0, 0.2

    f_dust = 0.0

    lines_width = 300.0

    emission = True
    
  
  [[dustatt_modified_starburst]]

    E_BV_lines = 0.0, 0.1, 0.2, 0.3, 0.5

    E_BV_factor = 0.44

    uv_bump_width = 35.0
 
    uv_bump_amplitude = 0.0

    powerlaw_slope = 0.0

    Ext_law_emission_lines = 1

    Rv = 3.1

 
\end{lstlisting}

\section{Comparison between our sample and the Duarte Puertas et al. (2017) sample}\label{app:dp17}
\begin{figure*}[htb]
    \centering
    \includegraphics[width=0.9\textwidth]{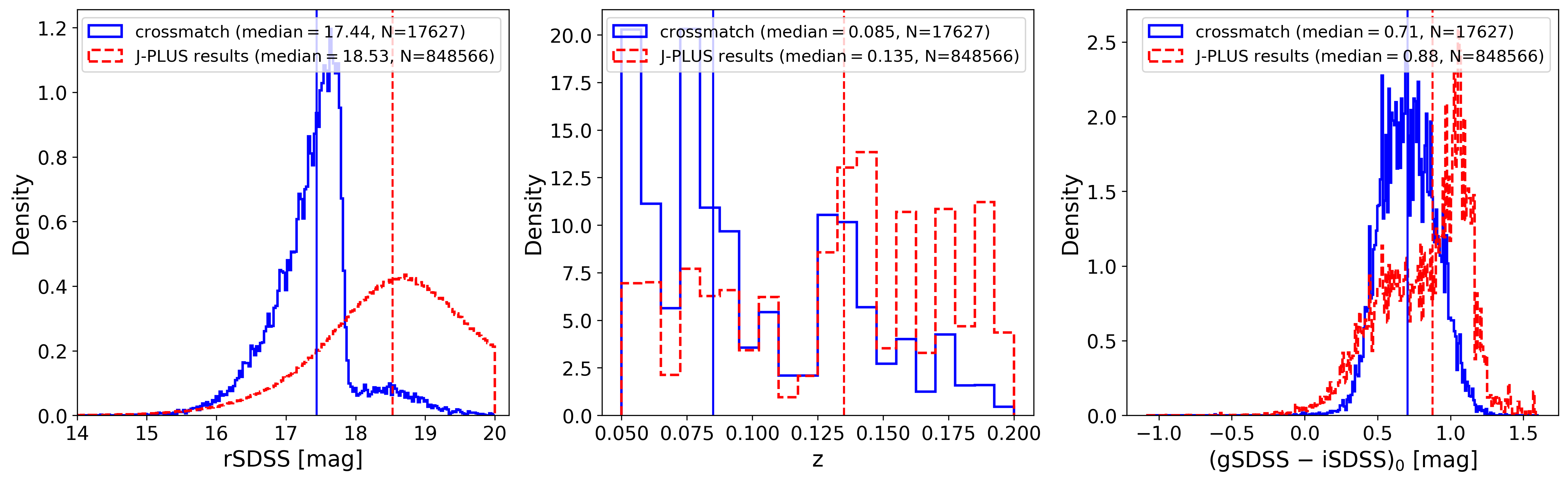}
    \caption{Distributions of the crossmatch sample (solid blue) and the full sample of our study (dashed red). {\it Left panel}: Distribution in $r$-band magnitude. {\it Middle panel}: Distribution of the redshift. {\it Right panel}: Distribution of the rest-frame color $(g-i)_0$.}
    \label{fig:DP_JPLUS_crossmatch}
\end{figure*}

In the first panel of Fig.~\ref{fig:DP_JPLUS_crossmatch}, the sample of \citet{puertas2017aperture} is centered on H$\alpha$, and is therefore more focused on blue galaxies. Their sample is also complete down to $r_{\rm SDSS} < 17.7$ mag, corresponding to the SDSS main galaxy sample, whereas our sample was selected with $r < 20$ mag. In the second panel of Fig.~\ref{fig:DP_JPLUS_crossmatch}, we compare the redshift distributions of the cross-matched sample between our study and \citet{puertas2017aperture}. The cross-matched sample contains more galaxies at lower redshift, whereas our full sample includes more galaxies at higher redshift, as reflected by the difference in the median values. In the third panel of Fig.~\ref{fig:DP_JPLUS_crossmatch}, the distribution of the rest-frame color $(g-i)$ is shown. We find that the cross-matched sample is more concentrated toward blue galaxies, whereas our sample shows a clearer bimodality.
\section{Single versus double Schechter model comparison}
\label{app:schechter_ic}
\begin{figure*}[t]
  \centering
  \includegraphics[width=\linewidth]{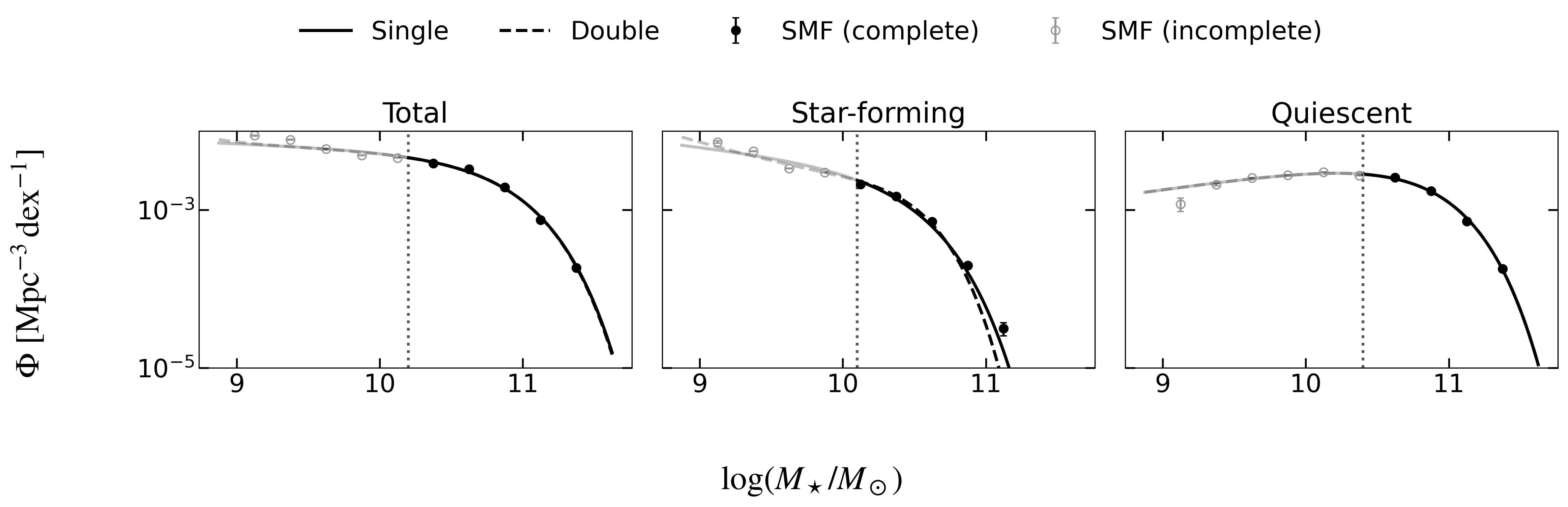}
  \caption{Comparison of best-fitting single- (solid) and double-Schechter (dashed) models over the observed SMFs (points with error bars). The curves show the Eddington-convolved (observed-space) model predictions from the \cite{obreschkow2018eddington} fitting framework. The vertical dotted lines mark the stellar-mass completeness limits at $z=0.2$ (from Fig.~\ref{fig:M_lim_2}); bins below these limits are shown as gray open symbols, and the model curves are faded below the completeness limit. For the total and SF-SMFs the double-Schechter curve corresponds to the standard (free) five-parameter optimization, while for the Q-SMF the double-Schechter curve corresponds to the stabilized two-stage (tight) procedure (Stage~B) described in the text.}
  \label{fig:schechter_single_double_overlay}
\end{figure*}

We fit Schechter functions with the \citet{obreschkow2018eddington} modified maximum-likelihood (MML) method implemented in \texttt{dftools::dffit}. In this approach the Schechter function is treated as a generative distribution in $\log M_\star$ and is convolved with the stellar-mass error distribution; selection effects are incorporated through the $1/V_{\max}$ weights described in Sect.~\ref{sec:vmax}. The single-Schechter function (SSF) has three free parameters $(\log_{10}\phi_1,\ \log_{10}M^\star,\ \alpha_1)$, while the double-Schechter function (DSF) adopts a shared characteristic mass $M^\star$ and adds a second component $(\log_{10}\phi_2,\ \alpha_2)$, for a total of five free parameters. For completeness, the best-fitting DSF parameters used in this comparison are reported in Table~\ref{tab:double_schechter_params_app}, and a direct visual comparison between the best-fitting SSF and DSF is shown in Fig.~\ref{fig:schechter_single_double_overlay}.

For each sample (total, SF, and Q) we obtain the best-fitting parameters with the MML method and compare the SSF and DSF through the Bayesian evidence returned by the fitting framework \citep{Kass1995,Trotta2008},
\begin{equation}
Z \equiv p(D\mid M) = \int p(D\mid \theta,M)\,p(\theta\mid M)\,d\theta,
\label{eq:evidence_appendix}
\end{equation}
where $D$ denotes the data, $M$ the model, $\theta$ the model parameters, $p(D\mid \theta,M)$ the likelihood, and $p(\theta\mid M)$ the prior. The Bayesian evidence thus quantifies the probability of the data under a given model after marginalizing over its parameter space, naturally accounting for both goodness of fit and model complexity. To ensure consistency with the interpretation of Fig.~\ref{fig:schechter_single_double_overlay}, this comparison is performed over the mass-complete interval for each population. We report the difference in log-evidence as $\Delta \ln Z \equiv \ln Z_{\rm DSF}-\ln Z_{\rm SSF}$, where negative values indicate that the SSF is preferred over the DSF. For the total, SF, and Q-SMFs, we obtain $\Delta \ln Z = -2611$, $-574$, and $-23$, respectively. In all three cases, the Bayesian evidence favors the SSF. As shown in Fig.~\ref{fig:schechter_single_double_overlay}, the best-fitting SSF and DSF curves are nearly indistinguishable over the fitted mass range.

The MML approach improves on a simple binned estimator by fitting the unbinned stellar-mass distribution directly, while incorporating the redshift-dependent selection function and the convolution with stellar-mass uncertainties, thus correcting for Eddington bias without introducing an arbitrary dependence on binning. In the present analysis, the SSF--DSF comparison is constrained mainly by the mass-complete regime at the upper redshift boundary, that is, above $M_{\rm lim}(z=0.2)$ for each population. The lower-mass bins are included in Fig.~\ref{fig:schechter_single_double_overlay} to illustrate the observed behavior of the SMF below the mass-complete regime, but they are not uniformly complete across $0.05\le z\le0.2$ and therefore do not constrain the SSF--DSF comparison.

\begin{figure}[!t]
  \centering
  \includegraphics[width=0.95\linewidth]{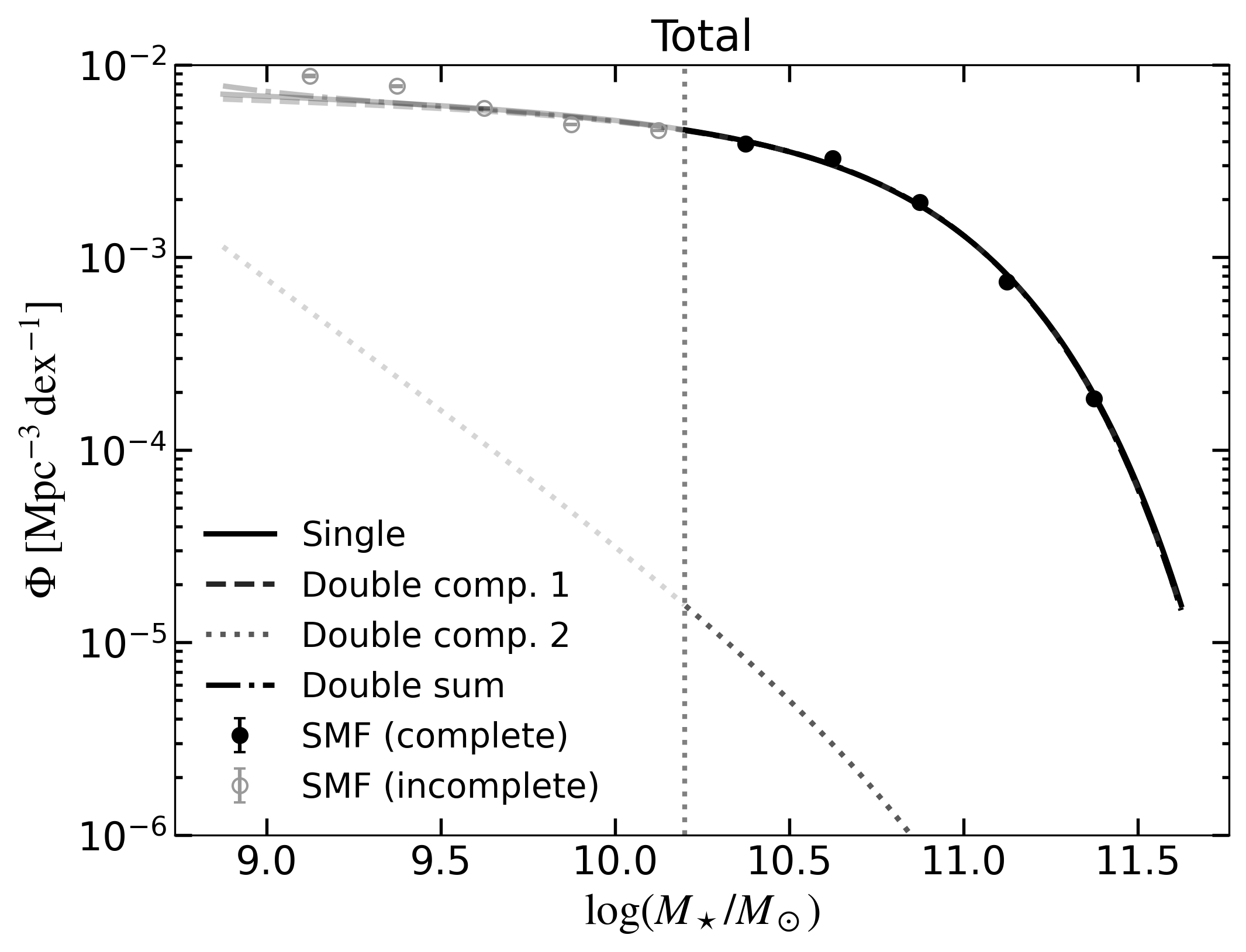}
  \caption{Double-Schechter decomposition for the total SMF. The observed SMF is shown with points and error bars. The two double-Schechter components are plotted separately (component 1: dashed; component 2: dotted), together with their sum (dash-dotted) and the best-fitting single-Schechter model (solid). The vertical dotted line marks the stellar-mass completeness limit at $z=0.2$; bins below this limit are shown as gray open symbols, and model curves are faded below the completeness limit.}
  \label{fig:dsf_total_components}
\end{figure}

The DSF fits used for Fig.~\ref{fig:schechter_single_double_overlay} are obtained as follows. For the total and SF-SMFs we use the standard single-stage DSF optimization (free fit), in which all five parameters are varied simultaneously within the MML framework. For the Q-SMF, the faint component is comparatively weakly constrained over the present mass-complete interval and exhibits strong degeneracies between $(\phi_2,\alpha_2,M^\star)$ when all parameters are left free. To ensure a well-posed DSF solution for Q galaxies, we therefore adopt a stabilized two-stage (tight) procedure: the bright end is first anchored with a high-mass prefit (Stage~A), and the full mass range is then refitted (Stage~B) using a pivot reparameterization of the faint component together with weakly informative priors and an ordering constraint ($\alpha_1>\alpha_2+0.3$). This strategy stabilizes the faint component without altering the bright-end description. For the total and Q-SMFs, however, the optimization drives the second component to a negligible normalization (Table~D.1), such that the resulting DSF is effectively indistinguishable from the SSF over the fitted mass range. This behavior is illustrated explicitly for the total SMF in Fig.~\ref{fig:dsf_total_components}, where the second component remains well below the dominant component and the summed DSF closely follows the SSF. We therefore interpret these cases as indicating that the second component is not meaningfully constrained by the present data, rather than as evidence against a DSF description of the local SMF in general.

Finally, we note that many low-redshift studies find that a DSF provides an improved description of the total SMF when extending to substantially lower stellar masses and/or when a wider dynamic range is available. In our case, the SSF--DSF comparison is effectively anchored to the mass-complete regime at the upper redshift boundary, $\log M_\star \ge M_{\rm lim}(z=0.2)$ (Fig.~\ref{fig:M_lim_2}), with $M_{\rm lim}(0.2)=10.2$ for the total sample, $10.1$ for SF galaxies, and $10.4$ for Q galaxies (vertical lines in Fig.~\ref{fig:schechter_single_double_overlay}). The low-mass bins displayed in Fig.~\ref{fig:schechter_single_double_overlay} are not complete across $0.05\le z\le0.2$. Within our adopted forward-modeling framework, which includes convolution with the stellar-mass uncertainties, the data in the mass-complete range do not statistically require an additional Schechter component, and the SSF is preferred.

\begin{table*}[t!]
\caption{Double-Schechter parameters for the three galaxy samples.}
\label{tab:double_schechter_params_app}
\centering
\renewcommand{\arraystretch}{1.3}
\begin{tabular}{lccccc}
\hline\hline
Sample &
$\log(\phi_{1,*})$ &
$\log(\phi_{2,*})$ &
$\log(M^{*}/M_{\odot})$ &
$\alpha_1$ &
$\alpha_2$ \\
 & [dex$^{-1}$\,Mpc$^{-3}$] & [dex$^{-1}$\,Mpc$^{-3}$] & & & \\
\hline
Q (tight; Stage~B) & $-2.632 \pm 0.002$ & $-6.11 \pm 0.43$ & $10.799 \pm 0.002$ & $-0.739 \pm 0.004$ & $-1.26 \pm 0.19$ \\
SF (free)          & $-2.787 \pm 0.008$ & $-3.15 \pm 0.02$ & $10.224 \pm 0.005$ & $-0.301 \pm 0.032$ & $-1.52 \pm 0.02$ \\
T (free)           & $-2.645 \pm 0.003$ & $-5.96 \pm 0.19$ & $10.863 \pm 0.002$ & $-1.055 \pm 0.003$ & $-2.34 \pm 0.09$ \\
\hline
\end{tabular}
\tablefoot{The total (T) and star-forming (SF) values correspond to the standard single-stage double-Schechter optimization (free fit). For Q galaxies we adopt the stabilized two-stage procedure (tight; Stage~B), which is the parameter set used for the Q double-Schechter curve shown in Fig.~\ref{fig:schechter_single_double_overlay}.}
\end{table*}
\FloatBarrier
\clearpage
\onecolumn
\section{Stellar mass function values}\label{app:smf}

Table~\ref{tab:log_smf_values_errors} lists the measured binned SMFs for the total, SF, and Q galaxy samples.

\begin{table}[h!]
\caption{Logarithmic SMF values for the total, SF, and Q galaxy samples.}
\label{tab:log_smf_values_errors}
\centering
\begin{tabular}{ccccc}
\hline\hline
Bin & $\log(M_\star/\mathrm{M}_\odot)$
    & $\log_{10}(\Phi_\mathrm{tot})$ [dex$^{-1}$\,Mpc$^{-3}$]
    & $\log_{10}(\Phi_\mathrm{SF})$ [dex$^{-1}$\,Mpc$^{-3}$]
    & $\log_{10}(\Phi_\mathrm{Q})$ [dex$^{-1}$\,Mpc$^{-3}$] \\
\hline
1  & 9.13  & $-2.0570 \pm 0.0069$ & $-2.140 \pm 0.019$ & $-2.93 \pm 0.18$ \\
2  & 9.37  & $-2.1101 \pm 0.0049$ & $-2.256 \pm 0.036$ & $-2.682 \pm 0.077$ \\
3  & 9.62  & $-2.2255 \pm 0.0039$ & $-2.474 \pm 0.036$ & $-2.595 \pm 0.056$ \\
4  & 9.87  & $-2.3089 \pm 0.0029$ & $-2.526 \pm 0.041$ & $-2.562 \pm 0.036$ \\
5  & 10.12 & $-2.3410 \pm 0.0018$ & $-2.674 \pm 0.045$ & $-2.520 \pm 0.036$ \\
6  & 10.37 & $-2.4112 \pm 0.0018$ & $-2.830 \pm 0.067$ & $-2.565 \pm 0.027$ \\
7  & 10.62 & $-2.4841 \pm 0.0019$ & $-3.149 \pm 0.092$ & $-2.592 \pm 0.025$ \\
8  & 10.87 & $-2.7122 \pm 0.0027$ & $-3.70 \pm 0.12$   & $-2.757 \pm 0.014$ \\
9  & 11.12 & $-3.1249 \pm 0.0045$ & $-4.50 \pm 0.17$   & $-3.1439 \pm 0.0073$ \\
10 & 11.37 & $-3.7352 \pm 0.0090$ & $-5.31 \pm 0.34$   & $-3.7471 \pm 0.0092$ \\
\hline
\end{tabular}
\tablefoot{These are the measured binned SMFs shown in Figs.~\ref{fig:smf_bins} and \ref{fig:schechter_single_double_overlay}. Units are dex$^{-1}$\,Mpc$^{-3}$. The quoted uncertainties are propagated logarithmic errors. Bins with $\log M_\star < M_{\rm lim}(z=0.2)$ are not complete over the full redshift range; however, the SMF is computed using the redshift-dependent completeness function $M_{\rm lim}(z)$ derived with the method of \citet{pozzetti2010}, shown in Fig.~\ref{fig:M_lim_2}.}
\end{table}

\section{Main headers of the catalog}

Table~\ref{tab:catalog_columns_vizier} summarizes the main columns of the released catalog together with their units and descriptions.

\begin{table}[h!]
\caption{Description of the released catalog columns.}
\label{tab:catalog_columns_vizier}
\centering
\begin{tabular}{lll}
\hline\hline
Column Name & Unit & Description \\
\hline
\texttt{ID}           & ---                     & Unique identifier: J-PLUS tile ID and object number \\
\texttt{RAJ2000}      & deg                     & Right Ascension (J2000) \\
\texttt{DEJ2000}      & deg                     & Declination (J2000) \\
\texttt{zphot}        & ---                     & Photometric redshift estimate (dimensionless) \\
\texttt{rSDSS}        & mag                     & MW extinction corrected apparent magnitude of J-PLUS r-band \\
\texttt{Mstar}        & log(M$_\odot$)          & Stellar mass estimated by \texttt{CIGALE} \\
\texttt{e\_Mstar}     & log(M$_\odot$)          & 1$\sigma$ uncertainty in stellar mass \\
\texttt{SFR}          & log(M$_\odot$ yr$^{-1}$) & SFR averaged over the last 10 Myr \\
\texttt{e\_SFR}       & log(M$_\odot$ yr$^{-1}$) & 1$\sigma$ uncertainty in SFR over 10 Myr \\
\texttt{g\_i}         & mag                     & Rest-frame (g$-$i) color \\
\texttt{e\_g\_i}      & mag                     & 1$\sigma$ uncertainty in rest frame (g$-$i) \\
\texttt{sglc\_prob\_star} & ---                 & SGLC star-galaxy probability \\
\texttt{ODDS}         & ---                     & BPZ ODDS parameter: redshift confidence \\
\texttt{Quiescent}    & ---                     & Flag: 1 if Q, 0 if SF \\
\hline
\end{tabular}
\tablefoot{The first five columns are based on J-PLUS DR3 photometric data, and the remaining columns correspond to derived physical properties obtained using \texttt{CIGALE}. Column names follow VizieR conventions.}
\end{table}

\end{document}